\documentclass[showpacs,amsmath,amssymb,aps,pra,
longbibliography,
10pt,reprint,superscriptaddress]{revtex4-1}
\usepackage{bm}
\usepackage[breaklinks=true,colorlinks=true,linkcolor=blue,urlcolor=blue,citecolor=blue]{hyperref}
\usepackage{dcolumn}
\usepackage{amsmath,amssymb}
\usepackage{mathdots}
\usepackage{natbib}
\usepackage{soul,color}
\usepackage{graphicx}
\usepackage{epstopdf}
\usepackage[note-name]{notes2bib}
\usepackage{mathptmx}
\usepackage{stmaryrd}

\begin{document}

\title{Vortex jets generated by edge defects in current-carrying superconductor thin strips}

\author{A.\,I. Bezuglyj}
    \affiliation{Institute for Theoretical Physics, NSC-KIPT, 61108 Kharkiv, Ukraine}
\author{V.\,A. Shklovskij}
    \affiliation{Physics Department, V. Karazin Kharkiv National University, 61022, Kharkiv, Ukraine}
\author{B.~Budinska}
    \affiliation{University of Vienna, Faculty of Physics, 1090 Vienna, Austria}
    \affiliation{University of Vienna, Vienna Doctoral School in Physics, 1090 Vienna, Austria}
\author{B.~Aichner}
    \affiliation{University of Vienna, Faculty of Physics, 1090 Vienna, Austria}
    \affiliation{University of Vienna, Vienna Doctoral School in Physics, 1090 Vienna, Austria}
\author{V.~M. Bevz}
    \affiliation{Physics Department, V. Karazin Kharkiv National University, 61022, Kharkiv, Ukraine}
\author{M.\,Yu.~Mikhailov}
    \affiliation{B. Verkin Institute for Low Temperature Physics and Engineering\\
    of the National Academy of Sciences of Ukraine, 61103, Kharkiv, Ukraine}
\author{D.\,Yu. Vodolazov}
    \affiliation{Institute for Physics of Microstructures, Russian Academy of Sciences,\\
    Nizhny Novgorod region 603087, Russia}
\author{W.~Lang}
    \affiliation{University of Vienna, Faculty of Physics, 1090 Vienna, Austria}
\author{O.\,V.~Dobrovolskiy}
    \email[Corresponding author: ]{oleksandr.dobrovolskiy@univie.ac.at}
    \affiliation{University of Vienna, Faculty of Physics, 1090 Vienna, Austria}
\date{\today}

\begin{abstract}
At sufficiently large transport currents $I_\mathrm{tr}$, a defect at the edge of a superconducting strip acts as a gate for the vortices entering into it. These vortices form a jet, which is narrow near the defect and expands due to the repulsion of vortices as they move to the opposite edge of the strip, giving rise to a transverse voltage $V_\perp$. Here, relying upon the equation of vortex motion under competing vortex-vortex and $I_\mathrm{tr}$-vortex interactions, we derive the vortex jet shapes in narrow ($\xi\ll w\lesssim\lambda_\mathrm{eff}$) and wide ($w\gg\lambda_\mathrm{eff}$) strips [$\xi$: coherence length, $w$: strip width, $\lambda_\mathrm{eff}$: effective penetration depth]. We predict a nonmonotonic dependence $V_\perp(I_\mathrm{tr})$ which can be measured with Hall voltage leads placed on the line $V_1V_2$ at a small distance $l$ apart from the edge defect and which changes its sign upon $l\shortrightarrow -l$ reversal. For narrow strips, we compare the theoretical predictions with experiment, by fitting the $V_\perp(I_\mathrm{tr},l)$ data for $1\,\mu$m-wide MoSi strips with single edge defects milled by a focused ion beam at distances $l = 16$-$80$\,nm from the line $V_1V_2$. For wide strips, the derived magnetic-field dependence of the vortex jet shape is in line with the recent experimental observations for vortices moving in Pb bridges with a narrowing. Our findings are augmented with the time-dependent Ginzburg-Landau simulations which reproduce the calculated vortex jet shapes and the $V_\perp(I_\mathrm{tr},l)$ maxima. Furthermore, with increase of $I_\mathrm{tr}$, the numerical modeling unveils the evolution of vortex jets to vortex rivers, complementing the analytical theory in the entire range of $I_\mathrm{tr}$.
\end{abstract}

\maketitle

\section{Introduction}\label{s1}
The recent great interest in superconductor thin strips with the critical current $I_\mathrm{c}$ approaching the depairing current $I_\mathrm{d}$ is caused by the required close-to-$I_\mathrm{d}$
bias regime of microstrip single-photon detectors\,\cite{Gol01apl,Nat12sst,Vod17pra,Kor20pra,Cha20apl}, ultra-fast vortex motion at large transport currents $I_\mathrm{tr}$\,\cite{Pui12pcs,Emb17nac,Vod19sst,Kog20prb,Pat21prb,Kog22prb}, and the phenomena of generation of sound\,\cite{Ivl99prb,Bul05prb} and spin waves\,\cite{Bes14prb,Dob21arx} at a few km/s vortex velocities. In this context, the issue of high $I_\mathrm{c}$ is related to blocking of the penetration of vortices via the strip edges\,\cite{Cle11prb,Vod12prb,Dob20nac}, its control via material and edge-barrier engineering\,\cite{Cer13njp,Loe19acs,Dob20nac}, and knowledge of the effects of various edge defects on the penetration and patterns of Abrikosov vortices\,\cite{Buz98pcs,Ala01pcs,Vod03pcs,Vod15sst,Siv18ltp,Bud22pra}.

The penetration of vortices into a superconductor is hampered by various types of surface and edge barriers\,\cite{Bra95rpp}, among which the Bean-Livingston\,\cite{Bea64prl} and the geometrical\,\cite{Zel94prl} barrier are most essential. The former arises due to the attraction of a vortex to its image at distances of the order of the London penetration depth $\lambda$ (in bulk superconductors) from the sample surface\,\cite{Bea64prl}. The geometrical barrier originates from the shape of the superconductor and it appears for samples different from an ellipsoid. In the case of superconductor thin strips with thickness $d\ll\lambda$, in which vortices interact mostly via the stray fields in the surrounding space, the vortex-vortex and vortex-image interaction length scale is determined by the noticeably larger effective penetration depth $\lambda_\mathrm{eff} = \lambda^2/d$\,\cite{Pea66jap,Kog94prb} while effects of the geometrical barrier are not essential\,\cite{Mik21prb}.

The patterns of moving vortices are determined by their mutual repulsion, interactions with the transport current and structural imperfections in the sample\,\cite{Bra95rpp}. For weak-pinning materials -- such as amorphous MoSi thin strips used in this work -- the effects of volume pinning are not essential. Then, if the Lorentz force $f_\mathrm{L} = \phi_0J_\mathrm{tr}/c$ [$\phi_0$: magnetic flux quantum] exerted on a vortex by the transport current of sheet density $J_\mathrm{tr}$ exceeds the force of attraction of the vortex to the sample edge, the edge barrier is suppressed. This suppression can be local in the case of a local increase of $J_\mathrm{tr}$ (current-crowding effect\,\cite{Ada13apl}), and it can be realized, e.\,g., in strips with an edge defect\,\cite{Buz98pcs,Ala01pcs,Vod03pcs,Cle11prb,Vod15sst,Dob20nac}. In this case, the defect acts as a gate\,\cite{Ala01pcs} for vortices entering into the superconductor strip and crossing it under the competing action of the Lorentz and vortex-vortex interaction forces. If the size of the defect is much larger than the coherence length $\xi$ in the superconductor, the defect can serve as a nucleation point for several vortex chains\,\cite{Emb17nac,Bud22pra}. Such chains form a vortex jet with the apex at the defect and expanding due to the repulsion of vortices as they move to the opposite film edge. If the defect size is $\backsimeq\xi$, the vortices will be entering into the strip consequentially. However, in the presence of fluctuations and inhomogeneities in the strip, the vortex chain will evolve into a diverging jet because of the intervortex repulsion.

The presence of a vortex velocity component along the superconducting strip gives rise to a \emph{transverse} voltage $V_\perp\neq0$, which is also known to appear in the different
physical contexts of vortex guiding\,\cite{Nie69jap,Sil03prb,Shk06prb,Rei08prb,Wor12prb,Dob19pra} and Hall\,\cite{Vin93prl,Lan01pcs,Wor09apl,Pui09prb,Zec18prb,Ric21sst} effects. However, vortex guiding results in an even-in-field transverse voltage $V_\perp(\bf B)$ while the Hall voltage is odd with respect to magnetic field $\bf B$ reversal. The appearance of a non-monotonic $V_\perp(I_\mathrm{tr})$ was also predicted for the annihilation of a vortex and antivortex entering into the strip via two displaced defects at its edges\,\cite{Gla86ltp} and confirmed experimentally for thin films with a cross-shaped geometry\,\cite{Ant91etp}. The essential differences of $V_\perp(I_\mathrm{tr})$ appearing because of the vortex jets in the present work is that it is (i) \emph{local}, i.\,e., it can only be measured with voltage leads placed rather close to the edge defect, (ii) appears also in \emph{zero} external magnetic field, (iii) \emph{changes its sign} with the change of the coordinate $l\shortrightarrow -l$ of the transverse voltage leads with respect to the edge defect, and (iv) appears because of the repulsion of \emph{several} (at least two) vortices.

Here, we predict theoretically and corroborate experimentally the appearance of the transverse voltage $V_\perp$ in the vicinity of an edge defect at zero external magnetic field. We use the dynamic equation for vortices moving under competing vortex-vortex and transport-current-vortex interaction conditions to derive analytically the transverse current-voltage curves $V_\perp(I_\mathrm{tr},l)$. The employed approach is justified at sufficiently large transport currents $I_\mathrm{tr}\geqslant I_\mathrm{c}$, i.\,e., when the edge barrier is suppressed by $I_\mathrm{tr}$. The major theoretical results are (i) the nonmonotonicity of $V_\perp$ as a function of $I_\mathrm{tr}$ and $l$, and (ii) analytical expressions for the vortex jet shapes in the cases of narrow ($\xi\ll w\lesssim\lambda_\mathrm{eff}$, $w$: strip width) and wide ($w\gg\lambda_\mathrm{eff}$) superconducting strips. Furthermore, we use a series of $1\,\mu$m-wide MoSi strips with artificially created edge defects (notches) milled by a focused ion beam (FIB) at different distances $l=16$-$80$\,nm from the transverse voltage leads to experimentally demonstrate the predicted non-monotonicity of $V_\perp(I_\mathrm{tr},l)$. In addition, we augment the analytical and experimental findings with the results of time-dependent Ginzburg-Landau (TDGL) simulations. The obtained vortex patterns reproduce qualitatively the calculated vortex jet shapes and the maxima in $V_\perp(I_\mathrm{tr},l)$ and illustrate the evolution of vortex jets to vortex rivers with increase of $I_\mathrm{tr}$, complementing the analytical theory in the entire range of transport currents.

The article is organized as follows. The vortex jet shapes and the transverse $I$-$V$ curves are derived in Secs.\,\ref{cNarrow} and \ref{cWide} for the cases of narrow and wide strips, respectively. The expression for the vortex velocity in a narrow strip is given in Sec.\,\ref{cVelocity} and the vortex jet shape in a wide strip in the presence of an external magnetic field is analyzed in Sec.\,\ref{cWideB}. The evolution of vortex jets to vortex rivers with increase of the transport current is discussed in Sec.\,\ref{cGL} relying upon the TDGL equation modeling. In Sec.\,\ref{cExp} we present the experimental data for thin MoSi strips, discuss them in comparison with the theoretical predictions and the TDGL simulations in Sec.\,\ref{cDiscussion}, and summarize the major obtained results in Sec.\,\ref{cConclusion}.

\section{Theory}

\subsection{Qualitative consideration}
\begin{figure}
    \includegraphics[width=0.85\linewidth]{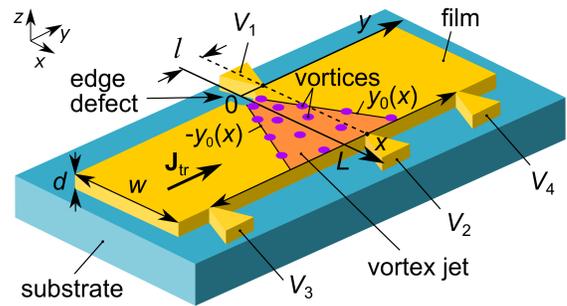}
    \caption{Geometry of the problem. The origin of coordinates is associated with an edge defect acting as a gate for vortices penetrating into the strip. The Lorentz force induced by the large transport current of density $\mathbf{J}_\mathrm{tr}$ makes the vortices to move to the opposite edge of the strip. The edges of the vortex jet in a narrow strip are determined by the equations $y=\pm y_0 (x)$. The transverse voltage $V_\perp=V_2-V_1$ associated with the crossing of the line $V_1V_2$ by vortices is measured with a pair of leads $V_1$ and $V_2$ located at a distance $l$ apart from the $x$-axis going through the center of the edge defect.}
    \label{f1}
\end{figure}

We first consider the problem qualitatively and then proceed to
its rigorous theoretical treatment. Our task is to elucidate the
appearance of the transverse voltage $V_\perp$ in a
superconducting strip with an edge defect, as schematically shown
in Fig.\,\ref{f1}. To this end, we consider the following scenario
of vortex penetration into the strip with an edge defect. At some
current $I_\mathrm{tr}=I_\mathrm{c}$ the current density reaches
the depairing current density $J_\mathrm{dep}$ near the defect,
the edge barrier vanishes and a vortex enters into the
superconductor. The supercurrent circulating around the vortex is
directed oppositely to $J_\mathrm{tr}$ near the defect, resulting
in a locally smaller $J_\mathrm{defect}< J_\mathrm{tr}$ and a
recovery of the edge barrier, thereby preventing the penetration
of other vortices. The recovery of the edge barrier is temporary
since the vortex moves toward the opposite edge of the strip due
to the transport current. The vortex motion is accompanied by a
redistribution of the transport current such that
$J_\mathrm{defect}$ reaches $J_\mathrm{dep}$ again and another
vortex enters into the strip.

In principle, in a pure uniform strip in the absence of a current
density gradient and/or fluctuations, the second vortex should
move along the same trajectory as the first vortex since the
first vortex does not create a perpendicular component of the
driving force acting on the second vortex. Moreover, the first
vortex creates a wake behind it, i.\,e., some track with a
suppressed superconducting order parameter, which attracts the
second vortex. The strength of the wake is determined by
nonequilibrium effects, the time-of-flight of the vortex across
the strip and, hence, by the magnitude of the transport current.

However, in the presence of fluctuations and inhomogeneities --
which cannot be avoided even in high-quality samples -- the second
vortex may escape from the wake and the vortex repulsion (which is
especially strong in narrow strips) will deflect the trajectory of
the second vortex away from the trajectory of the first vortex.
This deviation of the vortex trajectories away from the
transverse-to-strip direction ($x$-axis) will lead to the
appearance of a \emph{vortex jet}, as depicted in Fig.\,\ref{f1}.
Accordingly, when the edge defect is located at a distance $l
\gtrsim\xi$ from the line $V_1V_2$, the crossing of this line by
vortices leads to the appearance of a transverse voltage
$V_\perp\neq0$. In particular, when $I_\mathrm{tr}$ exceeds
$I_\mathrm{c}$ by only a small amount, the
vortex-interaction-induced longitudinal velocity component, $v_y$,
is relatively large in comparison with the Lorentz-force-induced
transverse velocity component $v_x$. In this regime,
$V_\perp(I_\mathrm{tr}) \propto I_\mathrm{tr}$ since the number $N$
of vortices crossing the line $V_1V_2$ per unit of time increases
as $N\propto I_\mathrm{tr}$. By contrast, the Lorentz-force-induced
perpendicular velocity component dominates the vortex dynamics at
$I_\mathrm{tr}\gg I_\mathrm{c}$. In this regime, for the line
$V_1V_2$ located not too close to the defect, though the velocity
of vortices is still increasing with increase of $I_\mathrm{tr}$,
the number of vortices crossing the line $V_1V_2$ is decreasing
faster. The competition between these two contributions leads to
the appearance of a maximum in $V_\perp(I_\mathrm{tr})$ and a
decrease of $V_\perp$ to zero with a further increase of
$I_\mathrm{tr}$.

In our hydrodynamic approximation with respect to the density of vortices and their velocity field, which allows for analytical expressions for the vortex jet shapes and the transverse $I$-$V$ curves, we will keep only the minimal number of essential terms which provide a non-monotonic dependence $V_\perp(I_\mathrm{tr})$. Namely, these are the vortex-vortex and transport-current-vortex interactions. In particular, we will neglect the interaction of vortices with the strip edges which could lead to quantitative corrections in the final expression, especially in the case of narrow strips.

\subsection{Vortex jet in a narrow strip}
\label{cNarrow}

In this subsection, we consider the case of a thin narrow
superconducting strip with thickness $d \ll \lambda$ and width
$\xi \ll w \lesssim \lambda_\mathrm{eff}$ carrying a transport
current $I_\mathrm{tr}$ in the absence of an external magnetic
field. The geometry of the problem is shown in Fig.\,\ref{f1}. The
origin of coordinates is associated with an edge defect acting as
a place of entry of vortices into the strip. We consider the
regime of sufficiently large transport currents $I_\mathrm{tr}
\geqslant I_\mathrm{c}$, where $I_\mathrm{c}$ is the current at
which the edge barrier is suppressed.

We assume that the force of interaction of a vortex located at the point $\bf r$ ($\xi \ll r \ll  \lambda_\mathrm{eff}$) with a vortex located at the origin of coordinates is given by the expression\,\cite{Pea64apl,Gen66boo}
\begin{equation}
    \label{1}
    {\bf F }({\bf r} )= \frac{\phi_{0}^2 }{8 \pi ^2 \lambda_\mathrm{eff}}\frac{\bf r}{r^2},
\end{equation}
where $\phi_{0}= hc/(2e)$ is the magnetic flux quantum. This force can be expressed as a gradient of the interaction potential\,\cite{Bla94rmp} $\varphi({\bf r} )= \displaystyle\frac{\phi_{0}^2}{8 \pi^2 \lambda_\mathrm{eff}} \ln{r/\lambda_\mathrm{eff}}$. For an ensemble of vortices located at points ${\bf r}_i$, the total interaction potential is given by
\begin{equation}
    \label{3}
    \varphi({\bf r} )=\sum_i  \frac{\phi_{0} ^2 }{8 \pi ^2 \lambda_\mathrm{eff}} \ln\frac{|{\bf r}- {\bf r}_i|}{\lambda_\mathrm{eff}}.
\end{equation}

With the two-dimensional vortex density $n({\bf r} )=\sum_i \delta_2 ({\bf r}- {\bf r}_i)$ the interaction potential can be expressed as
\begin{equation}
    \label{4}
    \nabla ^2\varphi({\bf r} )=  \frac{\phi_{0} ^2 }{4 \pi  \lambda_\mathrm{eff}}n({\bf r}).
\end{equation}

A stationary flow of vortices in the strip implies ${\rm div} (n {\bf v}) = 0$. The vortex trajectory can be found from the equation of the balance of forces acting on the vortex
\begin{equation}
    \label{5}
    \frac{\phi_{0}  }{c}\bigl[{\bf J}_\mathrm{tr} {\bf e}_{z}\bigr] +\nabla\varphi = \eta {\bf v},
\end{equation}
where ${\bf J}_\mathrm{tr}$ is the two-dimensional transport
current density, ${\bf e}_{z}$ the unit vector perpendicular to
the strip plane, $\eta$ the viscosity coefficient, and ${\bf v}$
the vortex velocity. Note that in the considered case of a narrow
strip with $w\lesssim\lambda_\mathrm{eff}$ the transport current
density is constant over the width of the strip. The width of the
vortex jet depends on the $x$ coordinate and is equal to
$2y_0(x)$.

We assume that the vortex density $n$ is constant over the jet
cross-section, that is $n=n(x,y)\equiv n(x)$. In this case, the
flux of vortices, which coincides with their frequency of
penetration into the strip, is given by
\begin{equation}
    \label{6}
    f_\mathrm{v}=2 y_0 (x) n(x)v_0 ,
\end{equation}
where $v_0=\phi_{0} J_\mathrm{tr}/c\eta$. In what follows we consider rather large transport currents which result in large vortex velocity components along the $x$-axis. This allows us to limit our consideration by the case of narrow jets with $\partial ^2\varphi/\partial  y^2\gg \partial ^2\varphi/\partial x^2$ and to simplify Eq.\,\eqref{4} as
\begin{equation}
    \label{6a}
    \frac{\partial ^2\varphi}{\partial  y^2}= \frac{\phi_{0} ^2 }{4 \pi  \lambda_\mathrm{eff}}\frac{f_\mathrm{v}}{2 y_0(x) v_0},
\end{equation}
from where
\begin{equation}
    \label{7}
    F_y \equiv\frac{\partial \varphi}{\partial  y}= \frac{\phi_{0} ^2 }{4 \pi \lambda_\mathrm{eff}}\frac{f_\mathrm{v} y}{2 y_0(x) v_0}.
\end{equation}

The equations of motion of the vortex $F_y= \eta \displaystyle\frac{dy}{dt}$ and $\displaystyle\frac{dx}{dt}=v_0$ yield the equation for the vortex trajectory
\begin{equation}
    \label{8}
    \frac{dy}{dx} = \frac{\phi_{0} ^2 f_\mathrm{v}}{8 \pi  \lambda_\mathrm{eff}\eta v_{0}^2}\frac{y}{ y_0 (x) }.
\end{equation}
The trajectories of the vortices at the edges of the jet are determined by the condition $y = y_0(x)$, from where one obtains
\begin{equation}
    \label{9}
    y_0 (x)= \frac{\phi_{0} ^2 f_\mathrm{v}}{8 \pi  \lambda_\mathrm{eff}\eta v_{0}^2}x \equiv \alpha_0 x,
\end{equation}
where $\alpha_0$ is the divergence angle of the jet. The integration constant is taken equal to zero since the size of the edge defect is assumed to be much smaller than the width of the vortex jet at $x = w$ (see Fig.\,\ref{f1}). It can be shown that the vortices inside the jet move along straight lines with slopes $|\alpha| < \alpha_0$. The above assumption $\partial ^2\varphi/\partial  y^2\gg \partial ^2\varphi/\partial x^2$ is justified in the case of a narrow vortex jet, that is, when $\alpha_0 \ll 1$.

Let the voltage leads $V_1$ and $V_2$ be located in such a way that their $y$-coordinates are equal to $l$, and $l$ is larger than the $y$-size of the edge defect, see Fig.\,\ref{f1}. Then, if the trajectories of vortices in the jet cross the dashed line $V_1V_2$, the transverse voltage $V_{\perp} \equiv V_2-V_1$ is induced between the leads $V_2$ and $V_1$. The appearance of the transverse voltage follows from the following considerations. The longitudinal voltage $V_\parallel \equiv V_4-V_3$ on the strip is determined by the average rate of change of the phase difference of the order parameter at its edges
\begin{equation}
    \label{9a}
    V =\frac{\hbar}{2e}\overline{\frac{d\phi}{dt}}.
\end{equation}

As one vortex passes through the strip, the phase difference of
the order parameter taken at two points located on both sides of
the row of moving vortices evolves by $2\pi$. Thus, the
longitudinal voltage $V_\parallel$ is proportional to the vortex
flux $V_\parallel = \pi\hbar f_\mathrm{v}/e$. Accordingly, the
transverse voltage $V_\perp$ is determined by the part of the flux
of vortices that crosses the line $V_1V_2$, namely,
$f_\mathrm{V_1V_2}=f_\mathrm{v}(\alpha_0-\alpha)/2\alpha_0$, where
$\alpha = l/w$ and $\alpha < \alpha_0$. For the calculation of $V_{\perp}(I_\mathrm{tr})$ we make use of the dependence $f_\mathrm{v} \propto (I_\mathrm{tr} - I_\mathrm{c})$ observed experimentally\,\cite{Emb17nac} for an edge defect in the form of a narrowing of the film. Using Eq.\,\eqref{9} for $\alpha_0$ and substituting $v_0=\phi_{0} I_\mathrm{tr}/w c\eta$, we obtain the sought-for expression
\begin{equation}
    \label{11}
    V_{\perp}(I_\mathrm{tr}) =
    \frac{\phi_0 k_\mathrm{I}}{2c}(I_\mathrm{tr} - I_\mathrm{c})
    \left (1-{\frac{8 \pi \lambda_\mathrm{eff} l I^2_\mathrm{tr}}{c^2 w^3 \eta k_\mathrm{I}(I_\mathrm{tr} - I_\mathrm{c})}}\right),
\end{equation}
where $k_\mathrm{I}$  is the coefficient of proportionality
between $f_\mathrm{v}$ and $(I_\mathrm{tr} - I_\mathrm{c})$ for
sufficiently large $I_\mathrm{tr}$.

\subsection{Vortex velocity in a narrow strip}
\label{cVelocity}

In recent years, thin strips in which vortices can move at high
velocities have attracted great
attention\,\cite{Emb17nac,Dob20nac,Leo20sst,Ust20sst,Kog20prb,Pat20prb,Pat21prb}.
This raises the problem of measuring the velocity of vortices
moving under the action of the Lorentz force. We note that in a
narrow strip, the $x$-component of the vortex velocity is
inversely proportional to the divergence angle of the jet
$\alpha_0$. In experimentally measured quantities, the expression
for the vortex velocity reads
\begin{equation}
    \label{17}
    v_0 = \frac{c ^2  d w V_\parallel}{8 \pi  \lambda^2 I_\mathrm{tr}\alpha_0},
\end{equation}
where $V_\parallel$ is the longitudinal voltage on the strip and $I_\mathrm{tr}$ is the current value at which $V_\perp$ vanishes.

The vortex velocity in a narrow strip can, therefore, be estimated from measurements of the transverse voltage $V_\perp$. Indeed, if a pair of voltage leads is located at a distance $l$ from the axis
$x$, then with an increase of the transport current, $V_\perp(I_\mathrm{tr})$ will become zero for a given $\alpha_0 = \alpha \equiv l/w$. By substituting $\alpha_0$ and $I_\mathrm{tr}|_{{V_\perp}=0}$ into Eq.\,\eqref{17} the velocity $v_0$ can be calculated.

\subsection{Vortex jet shape in a wide strip}
\label{cWide}

In this section, we consider the case of a thin wide
superconducting strip with thickness $d \ll \lambda$ and width
$w\gg\lambda_\mathrm{eff}$ carrying a transport current in the
absence of external magnetic field. Instead of re-deriving all the
expressions completely, we rather outline the modifications of the
results presented in Sec.\,\ref{cNarrow} for the case of wide
strips.

The distribution of the transport current density $J_\mathrm{tr}$
over the width of a wide strip has the well-known
form\,\cite{Rho62nat,Lar71etp}
\begin{equation}
    \label{10}
    J_\mathrm{tr}(x)= \frac{I_\mathrm{tr}}{\pi\sqrt{x(w-x)}},
\end{equation}
where $I_\mathrm{tr}$ is the total transport current flowing in
the strip. Equation\,\eqref{10} is inapplicable at distances of
the order of $\lambda_\mathrm{eff}$ from the strip boundaries.
However, this drawback is insignificant for the subsequent
analysis of strips with $w\gg\lambda_\mathrm{eff}$.

In the case of wide strips, the velocity $v_0$ in
Eqs.\,\eqref{6}--\eqref{8} is replaced by $v(x)$, where (not too
close to the strip edges)
\begin{equation}
    \label{12}
    v(x)= \frac{\phi_{0}I_\mathrm{tr}}{\pi c \eta \sqrt{x(w-x)}}.
\end{equation}

Since the $x$-component of the vortex velocity in wide strips
depends on the coordinate $x$, $dx/dt = v(x)$, one obtains the
following equation for the vortex trajectory
\begin{equation}
    \label{14}
    \frac{d y}{d x}= \frac{\phi_{0}^2}{8\pi \eta \lambda_\mathrm{eff}}\frac{f_\mathrm{v} y}{y_0 (x) v(x)^2}.
\end{equation}
As in the case of narrow strips, the jet boundary is determined by
the condition $y= y_0 (x)$, from where one obtains
\begin{equation}
    \label{15}
    y_0 (x)= \frac{\pi c ^2 \eta  f_\mathrm{v}}{8 \lambda_\mathrm{eff} I_\mathrm{tr}^2}\left(\frac{x^2 w}{2}-\frac{x^3}{3}\right).
\end{equation}

The shape of the vortex jet is largely determined by the Lorentz
force. Near the edges of a wide strip, where the current density
and, hence, the Lorentz force are relatively large, the curve
$y_0(x)$ has a low inclination. Near the center of the strip,
where the Lorentz force decreases, the dependence $y_0(x)$ becomes
steeper. Thus, in a wide strip, the diverging jet of vortices is
bounded by the curves with inflection points at $x = w/2$. The
vortex jet shape determined by Eq.\,\eqref{15} is in qualitative
agreement with the shape of the vortex patterns observed by
scanning SQUID-on-tip microscopy\,\cite{Emb17nac} in Pb bridges
with a narrowing, see Fig.\,\ref{f2}(a) and the inset in
Fig.\,3(b) in Ref.\,\cite{Emb17nac}.

Note that in the approach used it is hardly possible to obtain
perfect agreement between theory and experiment, since we use
Eq.\,\eqref{1} which quantitatively describes the interaction of
vortices in a narrow strip where the distance between vortices $r
\ll \lambda_\mathrm{eff}$. At the same time, the use of our
approach is not unreasonable for the description of wide strips.
An argument in its favor is the small width of the vortex jet,
since in a jet with a width of less than $\lambda_\mathrm{eff}$,
its expansion is mainly caused by the repulsion between vortices
located at distances less than $\lambda_\mathrm{eff}$. Equation
(\ref{14}) implies that the trajectories of vortices inside the
jet are given by the equality $ y (x)=\beta y_0 (x)$, where
parameter $|\beta| < 1$.

In the case of a wide strip, the transverse voltage is given by
\begin{equation}
    \label{16c}
    V_{\perp}(I_\mathrm{tr}) =
    \frac{\phi_0 k_\mathrm{I}}{2c}(I_\mathrm{tr} - I_\mathrm{c})
    \left (1-{\frac{48 \lambda_{\mathrm{eff}} I^2_\mathrm{tr}}{\pi c^2 w^3 \eta k_\mathrm{I}(I_\mathrm{tr}- I_\mathrm{c})}}\right).
\end{equation}
It is seen that $V_{\perp}(I_\mathrm{tr})$ has, again, a nonmonotonic behavior. At large currents $I_\mathrm{tr}$, such that $y_0(w) \le l$, $V_{\perp}$ is equal to zero.

\subsection{Vortex jet shape in a wide strip in a magnetic field}
\label{cWideB}

In this subsection we analyze the evolution of the vortex jet
shape in a wide strip in the presence of an external magnetic
field. Namely, when a magnetic field $\bf H$ is applied
perpendicular to the plane of a wide strip, a Meissner screening
current is induced in the superconductor, modifying the shape of
the vortex jet. In this subsection, for convenience, we assume
that the strip occupies the region $-1 < \tilde x <1$, where
$\tilde x$ is the dimensionless coordinate $\tilde x = 2x/w$.
Then, in the Meissner state, the current density has the
form\,\cite{Lar71etp}
\begin{equation}
    \label{16a}
    J(\tilde x)= \frac{4 I_\mathrm{tr} - c H w \tilde x}{2 \pi w (1-\tilde x^2)^{1/2}},
\end{equation}
where $H$ is the projection of the applied magnetic field onto the $z$-axis. If one introduces the dimensionless magnetic field $h = cwH/4I_\mathrm{tr}$, then the dependence $y_0(\tilde x)$ reads
\begin{equation}
    \label{16b}
    y_0 = \frac{\pi c^2 \eta  f_\mathrm{v} w^3}{64 \lambda_\mathrm{eff} I_{tr}^2} \left[\frac{2}{h^3}\ln \left(\frac{1+h}{ 1-h\tilde x}\right)-
    \frac{2(1+\tilde x)-h(1+\tilde x)^2}{h^2 (1-h\tilde x)}\right].
\end{equation}

\begin{figure}[t]
    \includegraphics[width=0.98\linewidth]{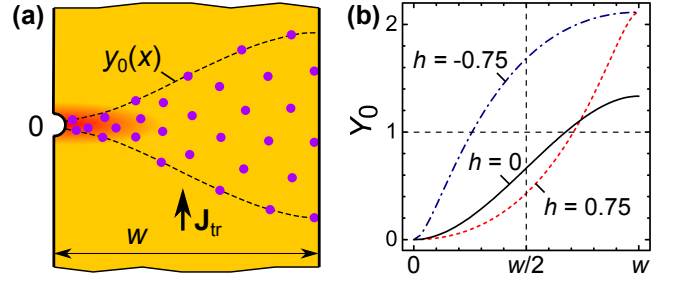}
    \caption{(a) Vortex jet shape in a wide strip.
    The jet edges are determined by Eq.\,\eqref{15}.
    (b) Coordinate dependence of the upper boundary of the vortex jet at three values of the reduced magnetic field $h$, as indicated. $Y_0$ is determined by the square bracket in Eq.\,\eqref{16b}.}
    \label{f2}
\end{figure}

Figure\,\ref{f2}(b) shows the coordinate dependence $y_0 (\tilde x)$ at three values of the reduced magnetic field $h$. Note that the shape of the vortex jet strongly depends on the polarity of the applied magnetic field. Indeed, at $h >0$, the direction of the screening current coincides with the direction of the transport current in the region $\tilde x <0$, which increases the Lorentz force acting on the vortex. By contrast, the same currents are oppositely directed at $\tilde x>0$, which leads to a decrease of the Lorentz force. When $h<0$, the situation changes to the opposite, since now the currents flow in the same direction when $\tilde x > 0$ and in the opposite directions when $\tilde x<0$.

\subsection{Vortex jet in the TDGL model}
\label{cGL}

The theoretical consideration in Secs.\,\ref{cNarrow}-\ref{cWideB}
is based on a hydrodynamic approach, wherein the vortex
distribution is described in terms of the vortex density. This
model allows for the analytical expressions for the vortex jet
shapes and the transverse $I$-$V$ curves. However, the
hydrodynamic approximation is connected with the unknown condition
for the vortex entry frequency via the edge defect. Besides, it is
difficult to take into account nonequilibrium effects associated
with the fast vortex motion at large transport currents.

\begin{figure*}[t]
    \includegraphics[width=0.98\linewidth]{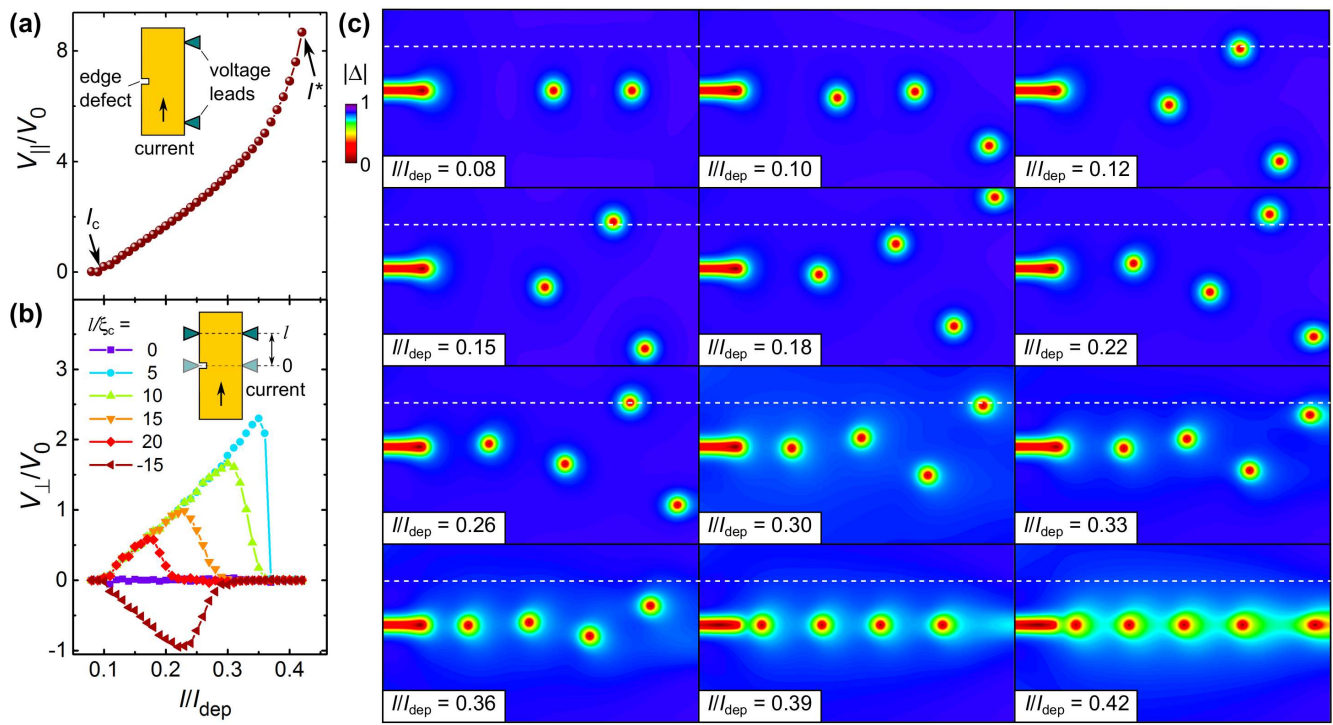}
    \caption{Longitudinal (a) and transverse (b) $I$-$V$ curves for a strip containing an edge defect, as calculated by the numerical solution of the TDGL equation.
    The voltage is in units of $V_0=k_\mathrm{B}T_\mathrm{c}/e$ and the current in units of the depairing current $I_\mathrm{dep}$ calculated in the dirty limit within the framework of the Usadel model\,\cite{Bud22pra}.
    (c) Snapshots of the modulus of the superconducting order parameter $|\Delta|(x,y)$ for a series of current values, as indicated, in the close-to-defect
    strip region with sizes $95\xi_\mathrm{c}\times 45\xi_\mathrm{c}$ ($x\times y$).
    Once the transport current $I_\mathrm{tr}$ exceeds the critical current $I_\mathrm{c}$, the vortices begin to enter via the edge defect (slit) and,
    with a further $I_\mathrm{tr}$ increase, form a divergent jet because of the repulsive interaction between them. At sufficiently large $I_\mathrm{tr}$,
    the opening angle of the jet is decreasing because of the $I_\mathrm{tr}$-vortex interaction dominating the vortex dynamics. When $I_\mathrm{tr}$ is
    approaching the instability current $I^\ast$, the vortex jets evolves to a vortex river, with the order parameter suppressed along the trajectory of
    the vortex motion. The horizontal dashed lines indicate the location of the transverse voltage leads with respect to the edge defect for $l /\xi_\mathrm{c} = 10$.}
    \label{f3}
\end{figure*}

As a complement to the analytical model introduced above, one can
take a discrete approach in which vortices enter the strip via the
defect one by one. This approach is based on the TDGL equation
which is solved in conjunction with the heat conductance equation.
Though no analytical expressions are available for the vortex jet
shapes and the transverse $I$-$V$ curves in this model, it takes
into account both, the condition for the vortex entry frequency
and nonequilibrium effects\,\cite{Bud22pra}. In addition, it
allows for analyzing the vortex patterns relying upon the
spatiotemporal evolution of the superconducting order parameter
$|\Delta|$.

Our model is valid at $T\approx T_\mathrm{c}$ and for short electron-electron scattering
times $\tau_\mathrm{ee}$. The latter condition implies that at any
time instant, electrons are thermalized among themselves and one
may introduce an electron temperature $T_\mathrm{e}$ differing
from the phonon temperature $T_\mathrm{ph}$ and the substrate
temperature $T_\mathrm{s}$. However, if $\tau_\mathrm{ee}$ is
comparable with the inelastic electron-phonon scattering time
$\tau_\mathrm{ep}$ (even if $\tau_\mathrm{ep}$ is reduced due to
the electron-electron interaction) the used model can give only qualitative predictions.

In the TDGL simulations, we consider a superconducting strip with
dimensions $100\xi_\mathrm{c}\times 130\xi_\mathrm{c}$ ($x\times
y$) where $\xi_\mathrm{c}=\sqrt{\hbar D/k_\mathrm{B}T_\mathrm{c}}
= \sqrt{1.76}\xi(0) \approx 7.8$\,nm for MoSi thin strips. Here,
$D$ is the electron diffusion coefficient and $T_\mathrm{c}$ is
the superconducting transition temperature. We use the normal
metal-superconductor boundary conditions at the ends of the strip
(along the $y$-axis), which allow us to employ a simple method for
injecting the current into the superconductor, and the boundary
conditions with vacuum along the $x$-axis. Details on the
considered equations and the numerical procedure were reported
elsewhere\,\cite{Bud22pra}. The edge defect was modeled as a
region (slit) with a local suppression of $T_\mathrm{c}$. The size
of the defect was chosen as $15\xi_\mathrm{c}\times
2\xi_\mathrm{c}$ ($x\times y$), as shown in Fig.\,\ref{f3}.

Figure\,\ref{f3}(a) and (b) show the calculated
$V_{\parallel}(I_\mathrm{tr})$ and $V_\perp(I_\mathrm{tr})$ curves
for the superconducting strip in a weak magnetic field $B=0.005
B_\mathrm{c0}$, where $B_\mathrm{c0}=\phi_0/2\pi\xi_\mathrm{c}^2$.
This magnetic field is slightly below the field $B_\mathrm{stop}$,
at which there is one sparse row of vortices at
$I_\mathrm{tr}<I_\mathrm{c}$ for a strip of length $L=260 \xi_\mathrm{c}$ while for a
strip of length $L=130 \xi_\mathrm{c}$ there are two vortices, see
Fig.\,\ref{f3}. The physical meaning of $B_\mathrm{stop}$ is half
of the magnetic field magnitude at which the edge barrier for
vortex entry is suppressed at $I_\mathrm{tr} = 0$. In the
simulations, the weak magnetic field is applied to increase the
number of vortices in the sample in the resistive state and to
expand the range of currents ($I_\mathrm{c},I^\ast$) of the
low-resistive regime which precedes the transition of the strip to
the normal state. The application of this small magnetic field
allows us to illustrate better the main results.

The evolution of the vortex patterns with increase of the
transport current is presented in Fig.\,\ref{f3}(c). Namely, we
find that at $I_\mathrm{tr}>I_\mathrm{c}$ two vortex rays are
formed -- one is deflected toward positive $y$ values and another
one in the direction of negative $y$ values with respect to the
edge defect. With an increase of $I_\mathrm{tr}$ the points, where
vortices exit the strip at the opposite edge, are displaced toward
$y=0$ so that the two vortex rays approach each other. Once the
vortex ray stops passing through the line $V_1V_2$, $V_\perp$
decreases to zero. Overall, the most essential features of the
transverse $I$-$V$ curves in Fig.\,\ref{f3}(b) can be summarized
as follows: $V_\perp(I_\mathrm{tr})=0$ for $l =0$. The
$V_\perp(I_\mathrm{tr})$ curve exhibits a maximum in the current
range $(I_\mathrm{c},I^\ast)$. The magnitude of the maximum in
$V_\perp(I_\mathrm{tr})$ decreases with increase of $l$. The
voltage $V_\perp$ changes its sign upon $l \shortrightarrow -l$
reversal.

Qualitatively, the same results were obtained for smaller and zero
magnetic fields. The only quantitative difference is that the
number of vortices and the divergent angle between the vortex rays
is smaller than that shown in Fig.\,\ref{f3}. For instance, at
$B=0$ only two vortices (one in each ray) are in the strip, but
the shape of $V_{\perp}(I_\mathrm{tr})$ is similar to that shown
in Fig.\,\ref{f3}(b). By contrast, with increase of the magnetic
field, the number of vortices increases, which leads to a wide
vortex jet instead of just two vortex rays. In this regime, the
vortices may exit at different points on the opposite edge of the
strip and this regime is realized already at $B=0.007 B_0$. At
larger magnetic fields, when $I_\mathrm{tr}$ exceeds some further
threshold value, vortices may enter not only via the edge defect
but also at other points along the sample edge because of the
suppression of the edge barrier. This regime requires further
investigations which are beyond the scope of the present work.

\section{Experiment}
\label{cExp}
\subsection{Samples}
The theoretical predictions for narrow strips were examined for a
series of 15\,nm-thick $1\,\mu$m-wide MoSi strips differing by the
location of an artificially-created edge defect (notch) with
respect to the perpendicular voltage leads, see Fig.\,\ref{f1} for
the geometry. MoSi was chosen as an amorphous superconductor with
a high structural uniformity and a very weak intrinsic (volume)
pinning, as previously concluded from the structural
characterization of the strips by transmission electron microscopy
and the magnetic-field dependence of the critical
current\,\cite{Bud22pra}. The MoSi films were deposited by dc
magnetron co-sputtering of Mo and Si targets onto Si/SiO$_2$
wafers, on top of a $5$\,nm-thick Si buffer layer, and covered
with a $3$\,nm-thick Si layer to prevent the strip oxidation.

\begin{figure}
    \includegraphics[width=0.9\linewidth]{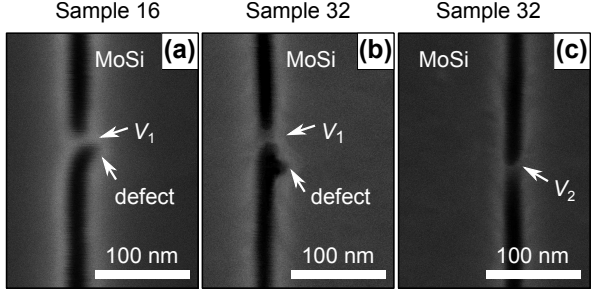}
    \caption{SEM images of the close-to-notch regions of (a) Sample 16 and (b) Sample 32.
    (c) Tilted-view SEM image of Sample 32 in the region of the transverse voltage lead at the opposite-to-notch edge.}
    \label{f4}
\end{figure}

For electrical resistance measurements the films were patterned
into six-probe geometries (Fig.\,\ref{f1}), with a strip length
$L=10\,\mu$m and width $w=1\,\mu$m. The voltage leads, with a
width of about $20$\,nm, were milled by FIB in a dual-beam
high-resolution scanning electron microscope (SEM: FEI Nova
NanoLab 600). FIB milling was done at $10$\,kV/$10$\,pA with a
pitch of $8$\,nm. In our previous study of the edge-barrier
effects on the vortex dynamics in wide MoSi
strips\,\cite{Bud22pra} we revealed that FIB milling allows for
the realization of very smooth strip edges. Specifically, the rms
edge roughness in the $y$-direction was less than $0.5$\,nm, as
deduced from the inspection of the strips by atomic force
microscopy over a distance of $500\,$nm along the edge. The
milling of the edges was accompanied by stopping of Ga ions within
a region of width $\sim10$\,nm along the edges, as inferred from
SRIM simulations and seen as lighter regions along the strip edges
in the SEM images in Fig.\,\ref{f4}.

The strips have a superconducting transition temperature
$T_\mathrm{c} = 6.43$\,K, resistivity $\rho_{\mathrm{7\,K}}
\approx148\,\mu\Omega$cm, upper critical field $B_\mathrm{c2}(0)
\approx 10.2$\,T, and $dB_{\mathrm{c}2}/dT =-2.23$\,T/K near
$T_\mathrm{c}$, yielding the electron diffusion coefficient
$D\approx0.49$\,cm$^2$/s, the coherence length $\xi(0) =
\sqrt{\hbar D /1.76k_\mathrm{B}T_\mathrm{c}} = 5.9$\,nm, the
penetration depth $\lambda(0) = 1.05\cdot10^{-3}
\sqrt{\rho_\mathrm{7K} /T_\mathrm{c}} \approx 495\,$nm, and
$\lambda_\mathrm{eff}(0) = \lambda^2(0)/d \approx 16.3\,\mu$m.
Thus, the investigated strips are thin and narrow, with $d \ll
\lambda$ and $\xi \ll w \lesssim \lambda_\mathrm{eff}$. With the
temperature dependence of the effective penetration
depth\,\cite{Tin04boo}, we make an estimate for
$\lambda_\mathrm{eff}(0.78T_\mathrm{c}) \approx 74\,\mu$m in our
experiments at $5\,$K. Hence, the large $\lambda_\mathrm{eff}$ in
conjunction with the weak intrinsic pinning ensures that the
dynamics of vortices entering via the edge defects is dominated by
the transport-current-vortex, vortex-edge and vortex-vortex
interactions. Altogether, this renders the MoSi strips as a
suitable experimental system for the examination of the
theoretical predictions of Sec.\,\ref{cNarrow}.

\subsection{Longitudinal $I$-$V$ curves}
\label{s5}

The $I$-$V$ curves were taken in the current-driven regime. The absolute value of the magnetic field in the vicinity of the sample was always smaller than $7\,\mu$T, as controlled by using a calibrated Hall sensor and representing zero-field conditions in our experiment. In what follows, we present the data for four samples, in which the middle of the notch was located at the distances $16$, $32$, $48$, and $80$\,nm from the line $V_1V_2$ between the transverse voltage leads. The samples are labeled as Sample 16, Sample 32, Sample 48, and Sample 80, corresponding to the parameter $\alpha\equiv l/w = 0.016$, $0.032$, $0.048$, and $0.080$, respectively. Also, an additional Sample A without an artificial defect was used for reference purposes. Due to the small distances between the notch and the voltage lead $V_1$ (close to the resolution limit for Ga FIB milling) multiple nominally identical structures were fabricated. SEM images of the close-to-notch regions of the strips chosen as Samples 16 and 32 for transport measurements are shown in Fig.\,\ref{f4}.

Figure\,\ref{f5}(a) presents the longitudinal $I$-$V$ curve for Sample A and Sample 48. The latter $I$-$V$ curve, within deviations of less than the symbol size, is representative for the $I$-$V$ curves of all samples containing a notch. Both $I$-$V$ curves in Fig.\,\ref{f5}(a) exhibit a zero-voltage plateau up to some critical current $I_\mathrm{c}$, which is $153\,\mu$A for Sample 48 and $265\,\mu$A for Sample A. The  $I_\mathrm{c}$ values were determined by using the $0.1\,\mu$V voltage criterion, as illustrated in Fig.\,\ref{f5}(a). Suppression of the edge barrier at $I_\mathrm{c}$ enables the penetration of vortices into the strips, resulting in rapid onsets of the low-resistive regime up to the abrupt jump to the highly-resistive state. These jumps occur in consequence of the flux-flow instability (FFI)\,\cite{Lar75etp,Lar86inb,Bez92pcs} at the current $I^\ast=261\pm3\,\mu$A for all samples with a notch. The instability current for Sample A is by about $15\%$ higher, $I^\ast_\mathrm{A}=302\,\mu$A, and the instability voltage $V^\ast_\mathrm{A}$ for Sample A is also higher by about $17\%$. These findings are in line with our previous observations for $1\,\mu$m-wide superconducting Nb-C strips with and without edge defects\,\cite{Dob20nac}.

\subsection{Critical current and its field dependence}

\begin{figure}
    \includegraphics[width=0.75\linewidth]{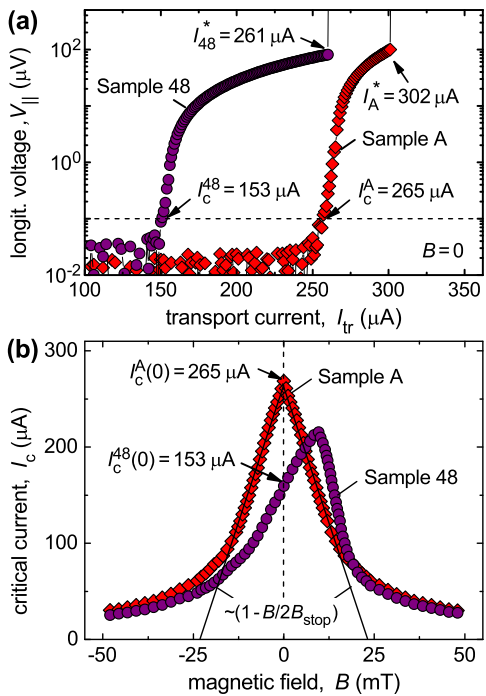}
    \caption{(a) Longitudinal $I$-$V$ curves for Sample 48 and Sample A in zero magnetic field.
    Dashed line indicates the $0.1\,\mu$V voltage criterion used for the determination of the critical current $I_\mathrm{c}$.
    (b) Symbols: Field dependence of the critical current $I_\mathrm{c}(B)$ for Sample 48 and Sample A. Solid lines:
    fits to the expression $I_\mathrm{c}(B) = I_\mathrm{c}(0)(1-B/2B_\mathrm{stop}$) with $B_\mathrm{stop} = 12$\,mT and $I_\mathrm{c}^\mathrm{A}(0)= 265\,\mu$A for Sample A.
    In all panels $T=5$\,K.}
    \label{f5}
\end{figure}

Figure\,\ref{f5}(b) shows the field dependences of the critical current for Sample A and Sample 48. The presence of a notch leads to a reduction of the critical current at $B =0$ and to a shift of the maximum in the originally symmetric $I_\mathrm{c}^A(B)$ under $\mathbf{B}$-reversal to about $+12$\,mT. At negative fields, the notch locally suppresses the edge barrier and thereby facilitates the entry of (anti)vortices. This leads to a small
reduction of $I^{48}_\mathrm{c}(B)$ up to $|B| \approx 50$\,mT at which the role of the volume pinning increases. At positive fields, when the vortices enter the microstrip from the opposite side, the notch does not affect the vortex entry and this is why $I^{48}_\mathrm{c}(B)$ is not affected by the presence of the notch at $B \gtrsim 15$\,mT.

The linear dependence $I_\mathrm{c}^{48}(B)$ at $B\shortrightarrow 0$ is indicative for a vortex-free (Meissner) state. Hence, with increase of $I_\mathrm{tr}$ the penetration of vortices is controlled by the locally suppressed edge barrier at the notch. In particular, for Sample A, $I_\mathrm{c}^\mathrm{A}(B)$ fits the dependence $I^\mathrm{A}_\mathrm{c}(B) = I^\mathrm{A}_\mathrm{c}(0\,\mathrm{T})(1- B/2B_\mathrm{stop})$ with $B_\mathrm{stop} = \phi_0/[2\sqrt3\pi\xi(T) w] \approx 12\,$mT, above which $I_\mathrm{c}^\mathrm{A}(B)$ decreases as $I_\mathrm{c}^\mathrm{A}(B) \propto B^{-1}$. We note that for our microstrip a current $I_\mathrm{tr}\simeq265\,\mu$A induces a self-field $B_\mathrm{self} =  0.5\mu_0 I_\mathrm{tr} w^{-1} \ln(2w/d) \simeq 0.8$\,mT which is much smaller than $B_\mathrm{stop}$ and, hence, the contribution of possible self-field effects to the observed crossover in $I_\mathrm{c}(B)$ at $B \approx12$\,mT is negligibly small.

With the temperature dependence of the depairing current $I_\mathrm{dep}(T) = I_\mathrm{dep}(0) (1 - (T/T_\mathrm{c})^{2})^{3/2}$, where $I_\mathrm{dep}(0) = 0.74 w [\Delta(0)]^{3/2}/(e R_\square \hbar D)$ for dirty superconductors\,\cite{Rom82prb,Cle12prb,Bud22pra} [$\Delta(0)$: superconducting gap at zero temperature, $R_\square$: sheet resistance] and the BCS ratio $\Delta(0) \approx 1.76 k_\mathrm{B}T_\mathrm{c}$ we obtain $I_\mathrm{dep}(0)\approx 1.27\,$mA and $I_\mathrm{dep}(\mathrm{5\,K})\approx 316\,\mu$A for our MoSi strips. This yields $I^{48}_\mathrm{c}/I_\mathrm{dep}\approx 0.5$, attesting to a strong local reduction of the barrier for vortex entry in the strips with an edge defect in comparison with Sample A, for which $I^\mathrm{A}_\mathrm{c}=265\,\mu$A yields $I_\mathrm{c}/I_\mathrm{dep}\approx 0.84$. Sample A has likely intrinsic edge defects, because its critical current is smaller than the theoretically expected depairing current. Moreover, we expect that the largest single defect suppresses $I^\ast$ as well since $V^\ast_\mathrm{A}$ is close to $V^\ast_{48}$. This is representative for all samples with notches, for which we assume the simultaneous motion of several vortices across the strip. Importantly, for all samples with a notch, the instability current $I^\ast = 261\pm3\,\mu$A is smaller than the critical current $I_\mathrm{c}^A = 265\,\mu$A for Sample A. This ensures that in the entire range of currents of our interest ($150$-$250\,\mu$A) the penetration of vortices occurs via the notch, as considered in the theoretical model.

\subsection{Transverse $I$-$V$ curves}

The transverse $I$-$V$ curves for all samples are presented in Fig.\,\ref{f6}(c). The transverse voltage $V_\perp$ is zero in the same range of currents $I_\mathrm{tr}\lesssim I_\mathrm{c}$ as $V_\parallel$ in Fig.\,\ref{f6}(a), and it exhibits a maximum in the regime of linear dependence $V_\parallel(I_\mathrm{tr})$. The maximum magnitude of $V_\perp$ is by about a factor of two smaller than the magnitude of $V_\parallel$. The $V_\perp(I_\mathrm{tr})$ curves for Samples 80, 48, and 32 exhibit a zero plateau above some threshold current $I_\alpha$ (indicated in Fig.\,\ref{f6}(c)), which shifts toward larger currents with decrease of $\alpha\equiv l/w$. For Sample 16, the dependence $V_\perp(I_\mathrm{tr})$ attains a maximum followed by a jump of $V_\perp$ to $55\,\mu$V upon the transition of the sample to the highly-resistive state.

Figure\,\ref{f6}(d) presents the transverse $I$-$V$ curve for sample A. While for all samples with an edge defect $V_\perp(I_\mathrm{tr})$ exhibit a maximum in the regime of linear dependence $V_\parallel(I_\mathrm{tr})$, the transverse voltage for Sample A fluctuates around the instrumental noise level in our setup up until the sample transition to the highly-resistive state at $I^\ast$. Accordingly, the occurrence of the maxima in $V_\perp(I_\mathrm{tr})$ can be clearly attributed to the presence of the milled notches rather close to the line $V_1V_2$ in Samples 16, 32, 48 and 80. The transverse voltage below $100$\,nV in Fig.\,\ref{f6}(d) also suggests that possible intrinsic edge defects are far away from the transverse potential leads in Sample A.

\section{Discussion}
\label{cDiscussion}
\subsection{Applicability of the model}

In Secs.\,\ref{cNarrow} and\,\ref{cWide} we
have analyzed the shape of vortex jets in narrow and wide strips and phenomena that arise due to the difference between the vortex jet and the vortex chain. It should be noted that the case of a superconducting strip in the Meissner state under consideration is different from the often studied case of vortex penetration into a strip that is in the critical state, where vortex avalanches (dendrites) are observed\,\cite{Ara01prl,Alt04rmp,Ara05prl,Bri16prb,Bri17prb,Sha19met,Jia20prb,Col21sst}. The considered regime is also different from the case of strips with rather strong intrinsic pinning at low magnetic fields, where the flux does not penetrate with a smooth advancing front, but instead as a series of irregularly shaped protrusions, resulting in $v^\ast\shortrightarrow0$ at $B\shortrightarrow0$\,\cite{Gri10prb}.

\begin{figure}
    \includegraphics[width=0.85\linewidth]{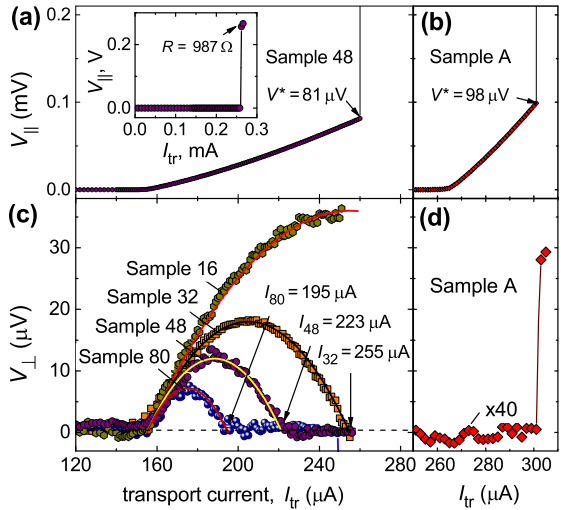}
    \caption{Longitudinal $I$-$V$ curves in the low-resistive regime for Sample 48 (a) and Sample A (b).
    Inset in (a): The same $I$-$V$ curve for Sample 48 with a transition to the normal state.
    (c) Transverse $I$-$V$ curves for all samples with an edge defect. Symbols: experiment; lines: fits to Eq.\,\eqref{11}.
    (d) Transverse $I$-$V$ curve for the reference Sample A without an edge defect.
    The voltage is multiplied with a factor of $40$. In all panels $T=5$\,K and $B =0$.}
    \label{f6}
\end{figure}

It is natural to assume that in the superconducting strip with one edge defect and zero applied magnetic field the vortices penetrate into the superconducting strip sequentially one after another via defect and form a vortex chain moving from one edge of the strip to another. Indeed, vortices move perpendicular to the edge of the strip due to transport current and there is no perpendicular component of the force (along the strip) due to other vortices (all vortices move along the same straight line connecting opposite edges of the strip). However, in the presence of fluctuations or inhomogeneities, this regime is unstable at \emph{not very large} velocities, i.\,e., when there is no channel with a suppressed order parameter, which appears either due to a finite relaxation time of the superconducting order parameter or due to heating. Due to the vortex-vortex repulsion, even small deviations of the trajectory of one vortex from the straight line lead to an instability of the whole vortex chain and the formation of a vortex jet.

Previously, it was shown that in the case of vortex flow in a narrow region across the superconducting strip (vortex river), the effects of self-heating play an important role\,\cite{Bez19prb}. When the vortices are injected through an edge defect, the heating of the vortex jet region leads to a decrease of the absolute value of the order parameter in this region and, therefore, to the lateral thermal pinning of the vortex jet. Such a thermal pinning justifies the assumption about the constancy of the vortex density in the cross-section of the jet.

Note that splitting of vortex chains was revealed in the past for \emph{fast-moving} vortices when one vortex river splits up into several rivers at a \emph{large current density gradient} across the bridge/strip width\,\cite{Emb17nac,Vod07prb}. By contrast, in the system considered in the present work, the current crowding near the small defect at the edge of a straight strip is localized near the defect and the current gradient near the edge has a small impact on the vortex dynamics. In particular, it was impossible to reproduce vortex jets in TDGL simulations (at zero or very weak magnetic fields) in the absence of fluctuations while vortex jets appeared in weak magnetic fields. In the latter case, the \emph{vortices already present in the strip} lead to the instability of the vortex chain. A further distinctive feature of our work from the wide bridges with narrowing considered in Ref.\,\cite{Emb17nac} is that in the regime of high vortex velocities, vortex jets \emph{appear} in Ref.\,\cite{Emb17nac}, whereas vortex jets \emph{vanish} evolving to vortex rivers in our work.

One of the conditions for the applicability of the analytical model is the smallness of the intervortex distances in comparison with the characteristic scales of the problem. This means that the vortex density must satisfy the following strong inequality: $n(w/2) \gg 1/y_{0}^2(w/2)$. Substitution of the density of vortices and the width of the jet into this inequality yields $f_{v}^2 \gg \phi_0 I_{tr}^3 \lambda_\mathrm{eff}/c^3 \eta ^2 w^4$. Thus, for the model to be valid, the frequency of penetration of vortices into the strip must be sufficiently high. At the same time, it turns out that at a current $I_\mathrm{tr} = 255\,\mu$A, due to the small width of the film, the average distance between the vortices is of the order of the coherence length. This means that the condition for the applicability of Eq.\,\eqref{1} is not satisfied, and only a qualitative description of the experiment on $1\,\mu$m-wide films can be expected from the presented model.

\subsection{Evaluation of the experiment}
In the theoretical model of Sec.\,\ref{cNarrow}, the edge-barrier effect on the vortex motion was neglected. This effect is most profound in narrow strips, leading to $I_\mathrm{c}\lesssim I_\mathrm{d}$ in high-quality samples. Phenomenologically, it can be accounted for by assuming that the average (over the strip width) vortex velocity $\langle v_0\rangle \propto f_\mathrm{v} \propto (I_\mathrm{tr} - I_\mathrm{c})$, as observed experimentally for an edge defect in the form of a narrowing of the film\,\cite{Emb17nac}. Indeed, the attraction of vortices to the edges can be described mathematically via the introduction of vortex images\,\cite{Bra95rpp}. We note that one can neglect the influence of the strip edges in Eq.\,\eqref{5} when $I_\mathrm{tr}\sim I_\mathrm{d}$, since the vortex-vortex interaction decays with the distance $r$ as $\sim 1/r$ already at $r>10 \xi$\,\cite{Bra95rpp}. However, the edge barrier strongly affects the time-of-flight $t$ of the vortex across the strip since near the edge where it enters the vortex moves very slowly and $t$ diverges when $I_\mathrm{tr}\rightarrow I_\mathrm{c}$ while $\langle v_0\rangle \rightarrow 0$.

Thus, we used the dependence $v_0 \propto (I_\mathrm{tr} - I_\mathrm{c})$ for fitting the experimental data (symbols in Fig.\,\ref{f6}(c)) to the expression $V_\perp = A (I_\mathrm{tr} - I_\mathrm{c}) - \alpha B (I_\mathrm{tr} - I_\mathrm{c})^2$ following from Eq.\,\eqref{11}, varying the coefficients $A$ and $B$ as two empirical parameters. The best fits were obtained for $A = (0.75\pm0.03)$\,V/A and $B = (225\pm5)$\,mV/mA$^2$.
The fits shown by solid lines in Fig.\,\ref{f6}(c) reproduce well the dome-shaped maxima in the experimental dependencies $V_\perp(I_\mathrm{tr})$ for all samples. Note that the agreement of the experimental $V_\perp(I_\mathrm{tr})$ with the parabolic dependence following from the theory indicates that, in our case, the edge defect injects an expanding jet of vortices rather than a chain of vortices.

Figure\,\ref{f6}(a) presents the longitudinal $I$-$V$ curve for Sample 48, emphasizing that the maximum in $V_\perp(I_\mathrm{tr})$ occurs in the quasi-linear regime of $V_\parallel(I_\mathrm{tr})$ and there is some range of currents ($223$-$255\,\mu$A) in which $V_\perp(I_\mathrm{tr})\approx0$ while $V_\parallel(I_\mathrm{tr})\propto (I_\mathrm{tr}-I_\mathrm{c})$. Thus, indeed, at sufficiently large $V_\parallel$ corresponding to sufficiently large vortex velocities $v_0$, the trajectories of vortices do not pass through the line $V_1V_2$. Finally, we note that the maximal magnitude of $V_\perp$ is decreasing with increase of the distance between the edge defect and the line $V_1V_2$. This attests to a local character of $V_\perp$, as predicted by the theory.

The $I_\alpha$ values in Fig.\,\ref{f6}(c) were also used to make estimates for the vortex velocity, as suggested theoretically in Sec.\,\ref{cVelocity}. Using Eq.\,\eqref{17}, in which $I_\mathrm{tr}$ is replaced by $(I_\mathrm{tr} - I_\mathrm{c}$), we obtain for $I_\mathrm{tr} = 255\,\mu$A the vortex velocity $v_0 \simeq100\,$km/s. This value is an order of magnitude greater than the maximal vortex velocities in Refs.\,\cite{Emb17nac,Bud22pra}. We attribute the overestimation of the vortex velocity to the specific dynamics of the order parameter. Namely, a rapidly moving vortex leaves behind a region with a suppressed order parameter for some time, and this region attracts subsequent vortices\,\cite{Gla86ltp}. As a result, the vortex jet narrows, and this narrowing is interpreted (based on Eq.\,\eqref{17}) as an increase in the vortex velocity.

The TDGL simulations suggest that in the broad range of transport currents $I_\mathrm{c} \lesssim I_\mathrm{tr} \lesssim I^\ast$ corresponding to $V_\perp\neq0$, the number of vortices moving in
the strip at each time instant is almost the same, see Fig.\,\ref{f3}. Herewith, the constancy of the average number of vortices in the strip at a given $I_\mathrm{tr}$ means that the penetration of vortices into the strip (at the frequency $f_\mathrm{v}$) occurs with the same frequency as the inverse of the time-of-flight $t = 1/f \equiv w/v_0$ needed for a vortex moving with velocity $v_0$ to cross the strip of width $w$\,\cite{Dob18apl}. Accordingly, if one assumes the motion of a single vortex and uses the Josephson relation $V_\parallel = \pi \hbar f_\mathrm{v}/e$, then the relation $v^\ast_0 = V^\ast_\parallel w /(\pi \hbar e)$ and the experimentally measured $V_\parallel^\ast = 81\,\mu$V at $I^\ast$ yield $v^\ast_0 \thickapprox40$\,km/s if $V_\parallel^\ast$ were associated with the motion of only one vortex. However, it is obvious that a vortex jet cannot be formed by one vortex. In particular, for our MoSi strips at $B=0$ the TDGL model predicts the presence of \emph{three vortices} as $I\shortrightarrow I^\ast$. However, this model also predicts $I^\ast\sim 0.6 I_\mathrm{dep}$ while in the experiment $I^\ast \sim I_\mathrm{dep}$. For three vortices moving in the strip, the instability velocity can be estimated as $v_0^\ast\approx 13$\,km/s.

There are two arguments in favor of this estimate: (i) Instability velocities $v^\ast$ of about  $13$\,km/s were recently deduced at $5$\,mT for $182\,\mu$m-wide MoSi thin strips which exhibit an
instability current of $I^\ast_{w=182\,\mu\mathrm{m}} \approx 50$\,mA\,\cite{Bud22pra}. At this field the number of vortices could be estimated via magnetic flux passing through the strip, from the standard relation $v^\ast = V^\ast_\parallel/(B c L)$. The scaling of the instability currents at zero field with the strip width within the framework of the edge-controlled FFI model\,\cite{Vod19sst,Dob20nac} suggests to expect $I^\ast_{w=1\,\mu\mathrm{m}} = (50/182)\,\mathrm{mA} = 275\,\mu$A, which is indeed rather close to $I^\ast=302\,\mu$A for the reference $1\,\mu$m-wide MoSi Sample A without an edge defect.

(ii) The vanish of $V_\perp$ at $I_\mathrm{tr} \gtrsim 255\,\mu$A for Sample 32 implies that, at close-to-instability currents, the angle of deviation of the vortex trajectories from the line $V_1V_2$ is less than $\arctan(0.032)\approx1.8^\circ$. This means that at such large currents the vortices move in a chain which then develops into a vortex river -- a chain of vortices with the depleted vortex cores because of the retarded
relaxation of quasiparticles outside the vortex cores. If one makes an estimate for the energy relaxation time $\tau_\epsilon$ of quasiparticles (normal electrons) left by fast-moving vortices, than three vortices in a vortex river, whose development into a normal domain mediates the onset of FFI, would have a separation $a$ of about $330$\,nm. In this case $\tau_\epsilon \approx a/v^\ast$ would yield $25$\,ps which is in line with
$\tau_\epsilon\approx 32$\,ps\,\cite{Bud22pra} deduced within the framework of the FFI theory\,\cite{Lar76etp,Lar86inb,Bez92pcs}, generalized by Doettinger \emph{et al}\,\cite{Doe95pcs}, for the $182\,\mu$m-wide strips made from the same MoSi strips.

\section{Conclusion}
\label{cConclusion} To sum up, we have predicted theoretically and
corroborated experimentally the appearance of the transverse
voltage $V_\perp$ in the vicinity of an edge defect in
superconducting strips at rather large transport currents in zero
magnetic field. This voltage is local, i.\,e., it can be measured
with transverse voltage leads placed at a rather small distance
$l$ apart from the edge defect and it changes its sign upon
$l\shortrightarrow -l$ reversal. The physical origin of $V_\perp$
is related to the motion of vortices penetrating via the edge
defect into the superconducting strip and forming a diverging
vortex jet as they move to the opposite edge of the strip. Due to
the different distribution of the transport current over the width
of the strip, the shape of the jet of vortices in a wide strip
(Eq.\,\eqref{15} and Fig.\,\ref{f2}) is qualitatively different
from the shape of the vortex jet in a narrow strip (Eq.\,\eqref{9}
and Fig.\,\ref{f1}).

The developed analytical model relies upon the dynamic equation
for vortices moving under competing vortex-vortex and
transport-current-vortex interactions and it is justified at
sufficiently large transport currents when the edge barrier is
already suppressed. The major theoretical results obtained in the
present work are (i) the analytical expressions \eqref{9} and
\eqref{15} for the vortex jet shapes in narrow and wide
superconducting strips, respectively, and (ii) the transverse
$I$-$V$ curves $V_\perp(I_\mathrm{tr})$ for the cases of narrow
(Eq.\,\eqref{11}) and wide (Eq.\,\eqref{16c}) superconducting
strips. For wide strips, the derived vortex jet shape is in
qualitative agreement with the recently observed patterns of
fast-moving vortices in Pb bridges with a
narrowing\,\cite{Emb17nac}.

For narrow strips, the theoretical predictions were compared with experiment, by fitting the
$V_\perp(I_\mathrm{tr},l)$ data for $1\,\mu$m-wide MoSi strips
with artificially created edge defects (notches) milled by FIB at
different distances from the transverse voltage leads. The
analytical and experimental findings have been further augmented
with the results of TDGL simulations which reproduce qualitatively
the calculated vortex jet shapes and the maxima in the
$V_\perp(I_\mathrm{tr},l)$ curves. In addition, the TDGL equation
modeling results have allowed us to illustrate the evolution of
vortex jets to vortex rivers with increase of $I_\mathrm{tr}$,
complementing the analytical theory in the entire range of
transport currents.

\section*{Acknowledgment}

A.I.B. is grateful to L.N.\,Davydov for discussions. O.V.D. thanks M. Huth for providing access to the dual-beam microscope and to R. Sachser for support with the nanofabrication. V.A.S. and M.Yu.M. acknowledge the Wolfgang Pauli Institute (WPI) Vienna for Scholarships within the framework of the Pauli Ukraine Project. M.Yu.M. acknowledges the Scholarship from the Krzysztof Skubiszewski Foundation. B.B. and B.A. acknowledge financial support by the Vienna Doctoral School in Physics (VDSP). V.M.B. acknowledges the European Cooperation in Science and Technology (E-COST) for support via Grants E-COST-GRANT-CA16218-5759aa9b and E-COST-GRANT-CA16218-46e403c7.
D.Y.V. acknowledges state contract No. 0035-2019-0021. This research is funded in whole, or in part, by the Austrian Science Fund (FWF), Grant No.\,I\,4865-N. Support by E-COST via COST Actions CA16218 (NANOCOHYBRI) and CA19108 (HiSCALE) is gratefully acknowledged.


\begin{thebibliography}{75}%
\makeatletter
\providecommand \@ifxundefined [1]{%
 \@ifx{#1\undefined}
}%
\providecommand \@ifnum [1]{%
 \ifnum #1\expandafter \@firstoftwo
 \else \expandafter \@secondoftwo
 \fi
}%
\providecommand \@ifx [1]{%
 \ifx #1\expandafter \@firstoftwo
 \else \expandafter \@secondoftwo
 \fi
}%
\providecommand \natexlab [1]{#1}%
\providecommand \enquote  [1]{``#1''}%
\providecommand \bibnamefont  [1]{#1}%
\providecommand \bibfnamefont [1]{#1}%
\providecommand \citenamefont [1]{#1}%
\providecommand \href@noop [0]{\@secondoftwo}%
\providecommand \href [0]{\begingroup \@sanitize@url \@href}%
\providecommand \@href[1]{\@@startlink{#1}\@@href}%
\providecommand \@@href[1]{\endgroup#1\@@endlink}%
\providecommand \@sanitize@url [0]{\catcode `\\12\catcode `\$12\catcode
  `\&12\catcode `\#12\catcode `\^12\catcode `\_12\catcode `\%12\relax}%
\providecommand \@@startlink[1]{}%
\providecommand \@@endlink[0]{}%
\providecommand \url  [0]{\begingroup\@sanitize@url \@url }%
\providecommand \@url [1]{\endgroup\@href {#1}{\urlprefix }}%
\providecommand \urlprefix  [0]{URL }%
\providecommand \Eprint [0]{\href }%
\providecommand \doibase [0]{http://dx.doi.org/}%
\providecommand \selectlanguage [0]{\@gobble}%
\providecommand \bibinfo  [0]{\@secondoftwo}%
\providecommand \bibfield  [0]{\@secondoftwo}%
\providecommand \translation [1]{[#1]}%
\providecommand \BibitemOpen [0]{}%
\providecommand \bibitemStop [0]{}%
\providecommand \bibitemNoStop [0]{.\EOS\space}%
\providecommand \EOS [0]{\spacefactor3000\relax}%
\providecommand \BibitemShut  [1]{\csname bibitem#1\endcsname}%
\let\auto@bib@innerbib\@empty
\bibitem [{\citenamefont {Gol'tsman}\ \emph {et~al.}(2001)\citenamefont
  {Gol'tsman}, \citenamefont {Okunev}, \citenamefont {Chulkova}, \citenamefont
  {Lipatov}, \citenamefont {Semenov}, \citenamefont {Smirnov}, \citenamefont
  {Voronov}, \citenamefont {Dzardanov}, \citenamefont {Williams},\ and\
  \citenamefont {Sobolewski}}]{Gol01apl}%
  \BibitemOpen
  \bibfield  {author} {\bibinfo {author} {\bibfnamefont {G.~N.}\ \bibnamefont
  {Gol'tsman}}, \bibinfo {author} {\bibfnamefont {O.}~\bibnamefont {Okunev}},
  \bibinfo {author} {\bibfnamefont {G.}~\bibnamefont {Chulkova}}, \bibinfo
  {author} {\bibfnamefont {A.}~\bibnamefont {Lipatov}}, \bibinfo {author}
  {\bibfnamefont {A.}~\bibnamefont {Semenov}}, \bibinfo {author} {\bibfnamefont
  {K.}~\bibnamefont {Smirnov}}, \bibinfo {author} {\bibfnamefont
  {B.}~\bibnamefont {Voronov}}, \bibinfo {author} {\bibfnamefont
  {A.}~\bibnamefont {Dzardanov}}, \bibinfo {author} {\bibfnamefont
  {C.}~\bibnamefont {Williams}}, \ and\ \bibinfo {author} {\bibfnamefont
  {R.}~\bibnamefont {Sobolewski}},\ }\bibfield  {title} {\enquote {\bibinfo
  {title} {Picosecond superconducting single-photon optical detector},}\ }\href
  {\doibase 10.1063/1.1388868} {\bibfield  {journal} {\bibinfo  {journal}
  {Appl. Phys. Lett.}\ }\textbf {\bibinfo {volume} {79}},\ \bibinfo {pages}
  {705--707} (\bibinfo {year} {2001})}\BibitemShut {NoStop}%
\bibitem [{\citenamefont {Natarajan}\ \emph {et~al.}(2012)\citenamefont
  {Natarajan}, \citenamefont {Tanner},\ and\ \citenamefont
  {Hadfield}}]{Nat12sst}%
  \BibitemOpen
  \bibfield  {author} {\bibinfo {author} {\bibfnamefont {Ch.~M.}\ \bibnamefont
  {Natarajan}}, \bibinfo {author} {\bibfnamefont {M.~G.}\ \bibnamefont
  {Tanner}}, \ and\ \bibinfo {author} {\bibfnamefont {R.~H.}\ \bibnamefont
  {Hadfield}},\ }\bibfield  {title} {\enquote {\bibinfo {title}
  {Superconducting nanowire single-photon detectors: physics and
  applications},}\ }\href {http://stacks.iop.org/0953-2048/25/i=6/a=063001}
  {\bibfield  {journal} {\bibinfo  {journal} {Supercond. Sci. Technol.}\
  }\textbf {\bibinfo {volume} {25}},\ \bibinfo {pages} {063001} (\bibinfo
  {year} {2012})}\BibitemShut {NoStop}%
\bibitem [{\citenamefont {Vodolazov}(2017)}]{Vod17pra}%
  \BibitemOpen
  \bibfield  {author} {\bibinfo {author} {\bibfnamefont {D.~Yu.}\ \bibnamefont
  {Vodolazov}},\ }\bibfield  {title} {\enquote {\bibinfo {title} {Single-photon
  detection by a dirty current-carrying superconducting strip based on the
  kinetic-equation approach},}\ }\href {\doibase
  10.1103/PhysRevApplied.7.034014} {\bibfield  {journal} {\bibinfo  {journal}
  {Phys. Rev. Appl.}\ }\textbf {\bibinfo {volume} {7}},\ \bibinfo {pages}
  {034014} (\bibinfo {year} {2017})}\BibitemShut {NoStop}%
\bibitem [{\citenamefont {Korneeva}\ \emph {et~al.}(2020)\citenamefont
  {Korneeva}, \citenamefont {Manova}, \citenamefont {Florya}, \citenamefont
  {Mikhailov}, \citenamefont {Dobrovolskiy}, \citenamefont {Korneev},\ and\
  \citenamefont {Vodolazov}}]{Kor20pra}%
  \BibitemOpen
  \bibfield  {author} {\bibinfo {author} {\bibfnamefont {Yu.~P.}\ \bibnamefont
  {Korneeva}}, \bibinfo {author} {\bibfnamefont {N.N.}\ \bibnamefont {Manova}},
  \bibinfo {author} {\bibfnamefont {I.N.}\ \bibnamefont {Florya}}, \bibinfo
  {author} {\bibfnamefont {M.~Yu.}\ \bibnamefont {Mikhailov}}, \bibinfo
  {author} {\bibfnamefont {O.V.}\ \bibnamefont {Dobrovolskiy}}, \bibinfo
  {author} {\bibfnamefont {A.A.}\ \bibnamefont {Korneev}}, \ and\ \bibinfo
  {author} {\bibfnamefont {D.~Yu.}\ \bibnamefont {Vodolazov}},\ }\bibfield
  {title} {\enquote {\bibinfo {title} {Different single-photon response of wide
  and narrow superconducting
  {${\mathrm{Mo}}_{x}{\mathrm{Si}}_{1\ensuremath{-}x}$} strips},}\ }\href
  {\doibase 10.1103/PhysRevApplied.13.024011} {\bibfield  {journal} {\bibinfo
  {journal} {Phys. Rev. Appl.}\ }\textbf {\bibinfo {volume} {13}},\ \bibinfo
  {pages} {024011} (\bibinfo {year} {2020})}\BibitemShut {NoStop}%
\bibitem [{\citenamefont {Charaev}\ \emph {et~al.}(2020)\citenamefont
  {Charaev}, \citenamefont {Morimoto}, \citenamefont {Dane}, \citenamefont
  {Agarwal}, \citenamefont {Colangelo},\ and\ \citenamefont
  {Berggren}}]{Cha20apl}%
  \BibitemOpen
  \bibfield  {author} {\bibinfo {author} {\bibfnamefont {I.}~\bibnamefont
  {Charaev}}, \bibinfo {author} {\bibfnamefont {Y.}~\bibnamefont {Morimoto}},
  \bibinfo {author} {\bibfnamefont {A.}~\bibnamefont {Dane}}, \bibinfo {author}
  {\bibfnamefont {A.}~\bibnamefont {Agarwal}}, \bibinfo {author} {\bibfnamefont
  {M.}~\bibnamefont {Colangelo}}, \ and\ \bibinfo {author} {\bibfnamefont
  {K.~K.}\ \bibnamefont {Berggren}},\ }\bibfield  {title} {\enquote {\bibinfo
  {title} {Large-area microwire {MoSi} single-photon detectors at 1550 nm
  wavelength},}\ }\href {\doibase 10.1063/5.0005439} {\bibfield  {journal}
  {\bibinfo  {journal} {Appl. Phys. Lett.}\ }\textbf {\bibinfo {volume}
  {116}},\ \bibinfo {pages} {242603} (\bibinfo {year} {2020})}\BibitemShut
  {NoStop}%
\bibitem [{\citenamefont {Puica}\ \emph {et~al.}(2012)\citenamefont {Puica},
  \citenamefont {Lang},\ and\ \citenamefont {Durrell}}]{Pui12pcs}%
  \BibitemOpen
  \bibfield  {author} {\bibinfo {author} {\bibfnamefont {I.}~\bibnamefont
  {Puica}}, \bibinfo {author} {\bibfnamefont {W.}~\bibnamefont {Lang}}, \ and\
  \bibinfo {author} {\bibfnamefont {J.~H.}\ \bibnamefont {Durrell}},\
  }\bibfield  {title} {\enquote {\bibinfo {title} {High velocity vortex
  channeling in vicinal {YBCO} thin films},}\ }\href
  {https://www.sciencedirect.com/science/article/pii/S0921453411005594}
  {\bibfield  {journal} {\bibinfo  {journal} {Physica C}\ }\textbf {\bibinfo
  {volume} {479}},\ \bibinfo {pages} {88--91} (\bibinfo {year}
  {2012})}\BibitemShut {NoStop}%
\bibitem [{\citenamefont {Embon}\ \emph {et~al.}(2017)\citenamefont {Embon},
  \citenamefont {Anahory}, \citenamefont {Jelic}, \citenamefont {Lachman},
  \citenamefont {Myasoedov}, \citenamefont {Huber}, \citenamefont {Mikitik},
  \citenamefont {Silhanek}, \citenamefont {Milosevic}, \citenamefont
  {Gurevich},\ and\ \citenamefont {Zeldov}}]{Emb17nac}%
  \BibitemOpen
  \bibfield  {author} {\bibinfo {author} {\bibfnamefont {L.}~\bibnamefont
  {Embon}}, \bibinfo {author} {\bibfnamefont {Y.}~\bibnamefont {Anahory}},
  \bibinfo {author} {\bibfnamefont {Z.~L.}\ \bibnamefont {Jelic}}, \bibinfo
  {author} {\bibfnamefont {E.~O.}\ \bibnamefont {Lachman}}, \bibinfo {author}
  {\bibfnamefont {Y.}~\bibnamefont {Myasoedov}}, \bibinfo {author}
  {\bibfnamefont {M.~E.}\ \bibnamefont {Huber}}, \bibinfo {author}
  {\bibfnamefont {G.~P.}\ \bibnamefont {Mikitik}}, \bibinfo {author}
  {\bibfnamefont {A.~V.}\ \bibnamefont {Silhanek}}, \bibinfo {author}
  {\bibfnamefont {M.~V.}\ \bibnamefont {Milosevic}}, \bibinfo {author}
  {\bibfnamefont {A.}~\bibnamefont {Gurevich}}, \ and\ \bibinfo {author}
  {\bibfnamefont {E.}~\bibnamefont {Zeldov}},\ }\bibfield  {title} {\enquote
  {\bibinfo {title} {Imaging of super-fast dynamics and flow instabilities of
  superconducting vortices},}\ }\href {\doibase 10.1038/s41467-017-00089-3}
  {\bibfield  {journal} {\bibinfo  {journal} {Nat. Commun.}\ }\textbf {\bibinfo
  {volume} {8}},\ \bibinfo {pages} {85} (\bibinfo {year} {2017})}\BibitemShut
  {NoStop}%
\bibitem [{\citenamefont {Vodolazov}(2019)}]{Vod19sst}%
  \BibitemOpen
  \bibfield  {author} {\bibinfo {author} {\bibfnamefont {D.~Yu.}\ \bibnamefont
  {Vodolazov}},\ }\bibfield  {title} {\enquote {\bibinfo {title} {Flux-flow
  instability in a strongly disordered superconducting strip with an edge
  barrier for vortex entry},}\ }\href {\doibase 10.1088/1361-6668/ab4168}
  {\bibfield  {journal} {\bibinfo  {journal} {Supercond. Sci. Technol.}\
  }\textbf {\bibinfo {volume} {32}},\ \bibinfo {pages} {115013} (\bibinfo
  {year} {2019})}\BibitemShut {NoStop}%
\bibitem [{\citenamefont {Kogan}\ and\ \citenamefont
  {Prozorov}(2020)}]{Kog20prb}%
  \BibitemOpen
  \bibfield  {author} {\bibinfo {author} {\bibfnamefont {V.~G.}\ \bibnamefont
  {Kogan}}\ and\ \bibinfo {author} {\bibfnamefont {R.}~\bibnamefont
  {Prozorov}},\ }\bibfield  {title} {\enquote {\bibinfo {title} {Interaction
  between moving {Abrikosov} vortices in {type-II} superconductors},}\ }\href
  {\doibase 10.1103/PhysRevB.102.024506} {\bibfield  {journal} {\bibinfo
  {journal} {Phys. Rev. B}\ }\textbf {\bibinfo {volume} {102}},\ \bibinfo
  {pages} {024506} (\bibinfo {year} {2020})}\BibitemShut {NoStop}%
\bibitem [{\citenamefont {Pathirana}\ and\ \citenamefont
  {Gurevich}(2021)}]{Pat21prb}%
  \BibitemOpen
  \bibfield  {author} {\bibinfo {author} {\bibfnamefont {W.~P. M.~R.}\
  \bibnamefont {Pathirana}}\ and\ \bibinfo {author} {\bibfnamefont
  {A.}~\bibnamefont {Gurevich}},\ }\bibfield  {title} {\enquote {\bibinfo
  {title} {Effect of random pinning on nonlinear dynamics and dissipation of a
  vortex driven by a strong microwave current},}\ }\href {\doibase
  10.1103/PhysRevB.103.184518} {\bibfield  {journal} {\bibinfo  {journal}
  {Phys. Rev. B}\ }\textbf {\bibinfo {volume} {103}},\ \bibinfo {pages}
  {184518} (\bibinfo {year} {2021})}\BibitemShut {NoStop}%
\bibitem [{\citenamefont {Kogan}\ and\ \citenamefont
  {Nakagawa}(2022)}]{Kog22prb}%
  \BibitemOpen
  \bibfield  {author} {\bibinfo {author} {\bibfnamefont {V.~G.}\ \bibnamefont
  {Kogan}}\ and\ \bibinfo {author} {\bibfnamefont {N.}~\bibnamefont
  {Nakagawa}},\ }\bibfield  {title} {\enquote {\bibinfo {title} {Dissipation of
  moving vortices in thin films},}\ }\href {\doibase
  10.1103/PhysRevB.105.L020507} {\bibfield  {journal} {\bibinfo  {journal}
  {Phys. Rev. B}\ }\textbf {\bibinfo {volume} {105}},\ \bibinfo {pages}
  {L020507} (\bibinfo {year} {2022})}\BibitemShut {NoStop}%
\bibitem [{\citenamefont {Ivlev}\ \emph {et~al.}(1999)\citenamefont {Ivlev},
  \citenamefont {Mej\'{\i}a-Rosales},\ and\ \citenamefont
  {Kunchur}}]{Ivl99prb}%
  \BibitemOpen
  \bibfield  {author} {\bibinfo {author} {\bibfnamefont {B.~I.}\ \bibnamefont
  {Ivlev}}, \bibinfo {author} {\bibfnamefont {S.}~\bibnamefont
  {Mej\'{\i}a-Rosales}}, \ and\ \bibinfo {author} {\bibfnamefont {M.~N.}\
  \bibnamefont {Kunchur}},\ }\bibfield  {title} {\enquote {\bibinfo {title}
  {Cherenkov resonances in vortex dissipation in superconductors},}\ }\href
  {\doibase 10.1103/PhysRevB.60.12419} {\bibfield  {journal} {\bibinfo
  {journal} {Phys. Rev. B}\ }\textbf {\bibinfo {volume} {60}},\ \bibinfo
  {pages} {12419--12423} (\bibinfo {year} {1999})}\BibitemShut {NoStop}%
\bibitem [{\citenamefont {Bulaevskii}\ and\ \citenamefont
  {Chudnovsky}(2005)}]{Bul05prb}%
  \BibitemOpen
  \bibfield  {author} {\bibinfo {author} {\bibfnamefont {L.~N.}\ \bibnamefont
  {Bulaevskii}}\ and\ \bibinfo {author} {\bibfnamefont {E.~M.}\ \bibnamefont
  {Chudnovsky}},\ }\bibfield  {title} {\enquote {\bibinfo {title} {Sound
  generation by the vortex flow in type-{II} superconductors},}\ }\href
  {\doibase 10.1103/PhysRevB.72.094518} {\bibfield  {journal} {\bibinfo
  {journal} {Phys. Rev. B}\ }\textbf {\bibinfo {volume} {72}},\ \bibinfo
  {pages} {094518} (\bibinfo {year} {2005})}\BibitemShut {NoStop}%
\bibitem [{\citenamefont {Bespalov}\ \emph {et~al.}(2014)\citenamefont
  {Bespalov}, \citenamefont {Mel'nikov},\ and\ \citenamefont
  {Buzdin}}]{Bes14prb}%
  \BibitemOpen
  \bibfield  {author} {\bibinfo {author} {\bibfnamefont {A.~A.}\ \bibnamefont
  {Bespalov}}, \bibinfo {author} {\bibfnamefont {A.~S.}\ \bibnamefont
  {Mel'nikov}}, \ and\ \bibinfo {author} {\bibfnamefont {A.~I.}\ \bibnamefont
  {Buzdin}},\ }\bibfield  {title} {\enquote {\bibinfo {title} {Magnon radiation
  by moving {Abrikosov} vortices in ferromagnetic superconductors and
  superconductor-ferromagnet multilayers},}\ }\href {\doibase
  10.1103/PhysRevB.89.054516} {\bibfield  {journal} {\bibinfo  {journal} {Phys.
  Rev. B}\ }\textbf {\bibinfo {volume} {89}},\ \bibinfo {pages} {054516}
  (\bibinfo {year} {2014})}\BibitemShut {NoStop}%
\bibitem [{\citenamefont {Dobrovolskiy}\ \emph {et~al.}(2021)\citenamefont
  {Dobrovolskiy}, \citenamefont {Wang}, \citenamefont {Vodolazov},
  \citenamefont {Budinska}, \citenamefont {Sachser}, \citenamefont {Chumak},
  \citenamefont {Huth},\ and\ \citenamefont {Buzdin}}]{Dob21arx}%
  \BibitemOpen
  \bibfield  {author} {\bibinfo {author} {\bibfnamefont {O.~V.}\ \bibnamefont
  {Dobrovolskiy}}, \bibinfo {author} {\bibfnamefont {Q.}~\bibnamefont {Wang}},
  \bibinfo {author} {\bibfnamefont {D.~Yu.}\ \bibnamefont {Vodolazov}},
  \bibinfo {author} {\bibfnamefont {B.}~\bibnamefont {Budinska}}, \bibinfo
  {author} {\bibfnamefont {R.}~\bibnamefont {Sachser}}, \bibinfo {author}
  {\bibfnamefont {A.V.}\ \bibnamefont {Chumak}}, \bibinfo {author}
  {\bibfnamefont {M.}~\bibnamefont {Huth}}, \ and\ \bibinfo {author}
  {\bibfnamefont {A.~I.}\ \bibnamefont {Buzdin}},\ }\bibfield  {title}
  {\enquote {\bibinfo {title} {Cherenkov radiation of spin waves by ultra-fast
  moving magnetic flux quanta},}\ }\href {https://arxiv.org/abs/2103.10156}
  {\bibfield  {journal} {\bibinfo  {journal} {arXiv:2103.10156}\ } (\bibinfo
  {year} {2021})}\BibitemShut {NoStop}%
\bibitem [{\citenamefont {Clem}\ and\ \citenamefont
  {Berggren}(2011)}]{Cle11prb}%
  \BibitemOpen
  \bibfield  {author} {\bibinfo {author} {\bibfnamefont {J.~R.}\ \bibnamefont
  {Clem}}\ and\ \bibinfo {author} {\bibfnamefont {K.~K.}\ \bibnamefont
  {Berggren}},\ }\bibfield  {title} {\enquote {\bibinfo {title}
  {Geometry-dependent critical currents in superconducting nanocircuits},}\
  }\href {\doibase 10.1103/PhysRevB.84.174510} {\bibfield  {journal} {\bibinfo
  {journal} {Phys. Rev. B}\ }\textbf {\bibinfo {volume} {84}},\ \bibinfo
  {pages} {174510} (\bibinfo {year} {2011})}\BibitemShut {NoStop}%
\bibitem [{\citenamefont {Vodolazov}(2012)}]{Vod12prb}%
  \BibitemOpen
  \bibfield  {author} {\bibinfo {author} {\bibfnamefont {D.~Y.}\ \bibnamefont
  {Vodolazov}},\ }\bibfield  {title} {\enquote {\bibinfo {title} {Saddle point
  states in two-dimensional superconducting films biased near the depairing
  current},}\ }\href {\doibase 10.1103/PhysRevB.85.174507} {\bibfield
  {journal} {\bibinfo  {journal} {Phys. Rev. B}\ }\textbf {\bibinfo {volume}
  {85}},\ \bibinfo {pages} {174507} (\bibinfo {year} {2012})}\BibitemShut
  {NoStop}%
\bibitem [{\citenamefont {Dobrovolskiy}\ \emph {et~al.}(2020)\citenamefont
  {Dobrovolskiy}, \citenamefont {Vodolazov}, \citenamefont {Porrati},
  \citenamefont {Sachser}, \citenamefont {Bevz}, \citenamefont {Mikhailov},
  \citenamefont {Chumak},\ and\ \citenamefont {Huth}}]{Dob20nac}%
  \BibitemOpen
  \bibfield  {author} {\bibinfo {author} {\bibfnamefont {O.~V.}\ \bibnamefont
  {Dobrovolskiy}}, \bibinfo {author} {\bibfnamefont {D.~Yu}\ \bibnamefont
  {Vodolazov}}, \bibinfo {author} {\bibfnamefont {F.}~\bibnamefont {Porrati}},
  \bibinfo {author} {\bibfnamefont {R.}~\bibnamefont {Sachser}}, \bibinfo
  {author} {\bibfnamefont {V.~M.}\ \bibnamefont {Bevz}}, \bibinfo {author}
  {\bibfnamefont {M.~Yu}\ \bibnamefont {Mikhailov}}, \bibinfo {author}
  {\bibfnamefont {A.~V.}\ \bibnamefont {Chumak}}, \ and\ \bibinfo {author}
  {\bibfnamefont {M.}~\bibnamefont {Huth}},\ }\bibfield  {title} {\enquote
  {\bibinfo {title} {Ultra-fast vortex motion in a direct-write {Nb-C}
  superconductor},}\ }\href {\doibase 10.1038/s41467-020-16987-y} {\bibfield
  {journal} {\bibinfo  {journal} {Nat. Commun.}\ }\textbf {\bibinfo {volume}
  {11}},\ \bibinfo {pages} {3291} (\bibinfo {year} {2020})}\BibitemShut
  {NoStop}%
\bibitem [{\citenamefont {Cerbu}\ \emph {et~al.}(2013)\citenamefont {Cerbu},
  \citenamefont {Gladilin}, \citenamefont {Cuppens}, \citenamefont {Fritzsche},
  \citenamefont {Tempere}, \citenamefont {Devreese}, \citenamefont
  {Moshchalkov}, \citenamefont {Silhanek},\ and\ \citenamefont {Van~de
  Vondel}}]{Cer13njp}%
  \BibitemOpen
  \bibfield  {author} {\bibinfo {author} {\bibfnamefont {D.}~\bibnamefont
  {Cerbu}}, \bibinfo {author} {\bibfnamefont {V.~N.}\ \bibnamefont {Gladilin}},
  \bibinfo {author} {\bibfnamefont {J.}~\bibnamefont {Cuppens}}, \bibinfo
  {author} {\bibfnamefont {J.}~\bibnamefont {Fritzsche}}, \bibinfo {author}
  {\bibfnamefont {J.}~\bibnamefont {Tempere}}, \bibinfo {author} {\bibfnamefont
  {J.~T.}\ \bibnamefont {Devreese}}, \bibinfo {author} {\bibfnamefont {V.~V.}\
  \bibnamefont {Moshchalkov}}, \bibinfo {author} {\bibfnamefont {A.~V.}\
  \bibnamefont {Silhanek}}, \ and\ \bibinfo {author} {\bibfnamefont
  {J.}~\bibnamefont {Van~de Vondel}},\ }\bibfield  {title} {\enquote {\bibinfo
  {title} {Vortex ratchet induced by controlled edge roughness},}\ }\href
  {http://stacks.iop.org/1367-2630/15/i=6/a=063022} {\bibfield  {journal}
  {\bibinfo  {journal} {New J. Phys.}\ }\textbf {\bibinfo {volume} {15}},\
  \bibinfo {pages} {063022--1--13} (\bibinfo {year} {2013})}\BibitemShut
  {NoStop}%
\bibitem [{\citenamefont {L{\"o}sch}\ \emph {et~al.}(2019)\citenamefont
  {L{\"o}sch}, \citenamefont {Alfonsov}, \citenamefont {Dobrovolskiy},
  \citenamefont {Keil}, \citenamefont {Engemaier}, \citenamefont {Baunack},
  \citenamefont {Li}, \citenamefont {Schmidt},\ and\ \citenamefont
  {B{\"u}rger}}]{Loe19acs}%
  \BibitemOpen
  \bibfield  {author} {\bibinfo {author} {\bibfnamefont {S.}~\bibnamefont
  {L{\"o}sch}}, \bibinfo {author} {\bibfnamefont {A.}~\bibnamefont {Alfonsov}},
  \bibinfo {author} {\bibfnamefont {O.~V.}\ \bibnamefont {Dobrovolskiy}},
  \bibinfo {author} {\bibfnamefont {R.}~\bibnamefont {Keil}}, \bibinfo {author}
  {\bibfnamefont {V.}~\bibnamefont {Engemaier}}, \bibinfo {author}
  {\bibfnamefont {S.}~\bibnamefont {Baunack}}, \bibinfo {author} {\bibfnamefont
  {G.}~\bibnamefont {Li}}, \bibinfo {author} {\bibfnamefont {O.~G.}\
  \bibnamefont {Schmidt}}, \ and\ \bibinfo {author} {\bibfnamefont
  {D.}~\bibnamefont {B{\"u}rger}},\ }\bibfield  {title} {\enquote {\bibinfo
  {title} {Microwave radiation detection with an ultra-thin free-standing
  superconducting niobium nanohelix},}\ }\href {\doibase
  10.1021/acsnano.8b07280} {\bibfield  {journal} {\bibinfo  {journal} {ACS
  Nano}\ }\textbf {\bibinfo {volume} {13}},\ \bibinfo {pages} {2948--2955}
  (\bibinfo {year} {2019})}\BibitemShut {NoStop}%
\bibitem [{\citenamefont {Buzdin}\ and\ \citenamefont
  {Daumens}(1998)}]{Buz98pcs}%
  \BibitemOpen
  \bibfield  {author} {\bibinfo {author} {\bibfnamefont {A.}~\bibnamefont
  {Buzdin}}\ and\ \bibinfo {author} {\bibfnamefont {M.}~\bibnamefont
  {Daumens}},\ }\bibfield  {title} {\enquote {\bibinfo {title} {Electromagnetic
  pinning of vortices on different types of defects},}\ }\href
  {https://www.sciencedirect.com/science/article/pii/S0921453497017292}
  {\bibfield  {journal} {\bibinfo  {journal} {Physica C}\ }\textbf {\bibinfo
  {volume} {294}},\ \bibinfo {pages} {257--269} (\bibinfo {year}
  {1998})}\BibitemShut {NoStop}%
\bibitem [{\citenamefont {Aladyshkin}\ \emph {et~al.}(2001)\citenamefont
  {Aladyshkin}, \citenamefont {Mel'nikov}, \citenamefont {Shereshevsky},\ and\
  \citenamefont {Tokman}}]{Ala01pcs}%
  \BibitemOpen
  \bibfield  {author} {\bibinfo {author} {\bibfnamefont {A.Yu.}\ \bibnamefont
  {Aladyshkin}}, \bibinfo {author} {\bibfnamefont {A.~S.}\ \bibnamefont
  {Mel'nikov}}, \bibinfo {author} {\bibfnamefont {I.~A.}\ \bibnamefont
  {Shereshevsky}}, \ and\ \bibinfo {author} {\bibfnamefont {I.~D.}\
  \bibnamefont {Tokman}},\ }\bibfield  {title} {\enquote {\bibinfo {title}
  {What is the best gate for vortex entry into {type-II} superconductor?}}\
  }\href {https://www.sciencedirect.com/science/article/pii/S092145340100288X}
  {\bibfield  {journal} {\bibinfo  {journal} {Physica C}\ }\textbf {\bibinfo
  {volume} {361}},\ \bibinfo {pages} {67--72} (\bibinfo {year}
  {2001})}\BibitemShut {NoStop}%
\bibitem [{\citenamefont {Vodolazov}\ \emph {et~al.}(2003)\citenamefont
  {Vodolazov}, \citenamefont {Maksimov},\ and\ \citenamefont
  {Brandt}}]{Vod03pcs}%
  \BibitemOpen
  \bibfield  {author} {\bibinfo {author} {\bibfnamefont {D.~Y.}\ \bibnamefont
  {Vodolazov}}, \bibinfo {author} {\bibfnamefont {I.~L.}\ \bibnamefont
  {Maksimov}}, \ and\ \bibinfo {author} {\bibfnamefont {E.~H.}\ \bibnamefont
  {Brandt}},\ }\bibfield  {title} {\enquote {\bibinfo {title} {Vortex entry
  conditions in {type-II} superconductors: {Effect} of surface defects},}\
  }\href {https://www.sciencedirect.com/science/article/pii/S0921453402018774}
  {\bibfield  {journal} {\bibinfo  {journal} {Physica C}\ }\textbf {\bibinfo
  {volume} {384}},\ \bibinfo {pages} {211--226} (\bibinfo {year}
  {2003})}\BibitemShut {NoStop}%
\bibitem [{\citenamefont {Vodolazov}\ \emph {et~al.}(2015)\citenamefont
  {Vodolazov}, \citenamefont {Ilin}, \citenamefont {Merker},\ and\
  \citenamefont {Siegel}}]{Vod15sst}%
  \BibitemOpen
  \bibfield  {author} {\bibinfo {author} {\bibfnamefont {D.~Yu.}\ \bibnamefont
  {Vodolazov}}, \bibinfo {author} {\bibfnamefont {K.}~\bibnamefont {Ilin}},
  \bibinfo {author} {\bibfnamefont {M.}~\bibnamefont {Merker}}, \ and\ \bibinfo
  {author} {\bibfnamefont {M.}~\bibnamefont {Siegel}},\ }\bibfield  {title}
  {\enquote {\bibinfo {title} {Defect-controlled vortex generation in
  current-carrying narrow superconducting strips},}\ }\href {\doibase
  10.1088/0953-2048/29/2/025002} {\bibfield  {journal} {\bibinfo  {journal}
  {Supercond. Sci. Technol.}\ }\textbf {\bibinfo {volume} {29}},\ \bibinfo
  {pages} {025002} (\bibinfo {year} {2015})}\BibitemShut {NoStop}%
\bibitem [{\citenamefont {Sivakov}\ \emph {et~al.}(2018)\citenamefont
  {Sivakov}, \citenamefont {Turutanov}, \citenamefont {Kolinko},\ and\
  \citenamefont {Pokhila}}]{Siv18ltp}%
  \BibitemOpen
  \bibfield  {author} {\bibinfo {author} {\bibfnamefont {A.~G.}\ \bibnamefont
  {Sivakov}}, \bibinfo {author} {\bibfnamefont {O.~G.}\ \bibnamefont
  {Turutanov}}, \bibinfo {author} {\bibfnamefont {A.~E.}\ \bibnamefont
  {Kolinko}}, \ and\ \bibinfo {author} {\bibfnamefont {A.~S.}\ \bibnamefont
  {Pokhila}},\ }\bibfield  {title} {\enquote {\bibinfo {title} {Spatial
  characterization of the edge barrier in wide superconducting films},}\ }\href
  {\doibase 10.1063/1.5024540} {\bibfield  {journal} {\bibinfo  {journal} {Low
  Temp. Phys.}\ }\textbf {\bibinfo {volume} {44}},\ \bibinfo {pages} {226--232}
  (\bibinfo {year} {2018})}\BibitemShut {NoStop}%
\bibitem [{\citenamefont {Budinsk\'a}\ \emph {et~al.}(2022)\citenamefont
  {Budinsk\'a}, \citenamefont {Aichner}, \citenamefont {Vodolazov},
  \citenamefont {Mikhailov}, \citenamefont {Porrati}, \citenamefont {Huth},
  \citenamefont {Chumak}, \citenamefont {Lang},\ and\ \citenamefont
  {Dobrovolskiy}}]{Bud22pra}%
  \BibitemOpen
  \bibfield  {author} {\bibinfo {author} {\bibfnamefont {B.}~\bibnamefont
  {Budinsk\'a}}, \bibinfo {author} {\bibfnamefont {B.}~\bibnamefont {Aichner}},
  \bibinfo {author} {\bibfnamefont {D.~Yu.}\ \bibnamefont {Vodolazov}},
  \bibinfo {author} {\bibfnamefont {M.~Yu.}\ \bibnamefont {Mikhailov}},
  \bibinfo {author} {\bibfnamefont {F.}~\bibnamefont {Porrati}}, \bibinfo
  {author} {\bibfnamefont {M.}~\bibnamefont {Huth}}, \bibinfo {author}
  {\bibfnamefont {A.V.}\ \bibnamefont {Chumak}}, \bibinfo {author}
  {\bibfnamefont {W.}~\bibnamefont {Lang}}, \ and\ \bibinfo {author}
  {\bibfnamefont {O.V.}\ \bibnamefont {Dobrovolskiy}},\ }\bibfield  {title}
  {\enquote {\bibinfo {title} {Rising speed limits for fluxons via edge-quality
  improvement in wide {MoSi} thin films},}\ }\href {\doibase
  10.1103/PhysRevApplied.17.034072} {\bibfield  {journal} {\bibinfo  {journal}
  {Phys. Rev. Appl.}\ }\textbf {\bibinfo {volume} {17}},\ \bibinfo {pages}
  {034072} (\bibinfo {year} {2022})}\BibitemShut {NoStop}%
\bibitem [{\citenamefont {Brandt}(1995)}]{Bra95rpp}%
  \BibitemOpen
  \bibfield  {author} {\bibinfo {author} {\bibfnamefont {E.~H.}\ \bibnamefont
  {Brandt}},\ }\bibfield  {title} {\enquote {\bibinfo {title} {The flux-line
  lattice in superconductors},}\ }\href
  {http://stacks.iop.org/0034-4885/58/i=11/a=003} {\bibfield  {journal}
  {\bibinfo  {journal} {Rep. Progr. Phys.}\ }\textbf {\bibinfo {volume} {58}},\
  \bibinfo {pages} {1465--1594} (\bibinfo {year} {1995})}\BibitemShut {NoStop}%
\bibitem [{\citenamefont {Bean}\ and\ \citenamefont
  {Livingston}(1964)}]{Bea64prl}%
  \BibitemOpen
  \bibfield  {author} {\bibinfo {author} {\bibfnamefont {C.~P.}\ \bibnamefont
  {Bean}}\ and\ \bibinfo {author} {\bibfnamefont {J.~D.}\ \bibnamefont
  {Livingston}},\ }\bibfield  {title} {\enquote {\bibinfo {title} {Surface
  barrier in type-{II} superconductors},}\ }\href {\doibase
  10.1103/PhysRevLett.12.14} {\bibfield  {journal} {\bibinfo  {journal} {Phys.
  Rev. Lett.}\ }\textbf {\bibinfo {volume} {12}},\ \bibinfo {pages} {14--16}
  (\bibinfo {year} {1964})}\BibitemShut {NoStop}%
\bibitem [{\citenamefont {Zeldov}\ \emph {et~al.}(1994)\citenamefont {Zeldov},
  \citenamefont {Larkin}, \citenamefont {Geshkenbein}, \citenamefont
  {Konczykowski}, \citenamefont {Majer}, \citenamefont {Khaykovich},
  \citenamefont {Vinokur},\ and\ \citenamefont {Shtrikman}}]{Zel94prl}%
  \BibitemOpen
  \bibfield  {author} {\bibinfo {author} {\bibfnamefont {E.}~\bibnamefont
  {Zeldov}}, \bibinfo {author} {\bibfnamefont {A.~I.}\ \bibnamefont {Larkin}},
  \bibinfo {author} {\bibfnamefont {V.~B.}\ \bibnamefont {Geshkenbein}},
  \bibinfo {author} {\bibfnamefont {M.}~\bibnamefont {Konczykowski}}, \bibinfo
  {author} {\bibfnamefont {D.}~\bibnamefont {Majer}}, \bibinfo {author}
  {\bibfnamefont {B.}~\bibnamefont {Khaykovich}}, \bibinfo {author}
  {\bibfnamefont {V.~M.}\ \bibnamefont {Vinokur}}, \ and\ \bibinfo {author}
  {\bibfnamefont {H.}~\bibnamefont {Shtrikman}},\ }\bibfield  {title} {\enquote
  {\bibinfo {title} {Geometrical barriers in high-temperature
  superconductors},}\ }\href {\doibase 10.1103/PhysRevLett.73.1428} {\bibfield
  {journal} {\bibinfo  {journal} {Phys. Rev. Lett.}\ }\textbf {\bibinfo
  {volume} {73}},\ \bibinfo {pages} {1428--1431} (\bibinfo {year}
  {1994})}\BibitemShut {NoStop}%
\bibitem [{\citenamefont {Pearl}(1966)}]{Pea66jap}%
  \BibitemOpen
  \bibfield  {author} {\bibinfo {author} {\bibfnamefont {J.}~\bibnamefont
  {Pearl}},\ }\bibfield  {title} {\enquote {\bibinfo {title} {Structure of
  superconductive vortices near a metal-air interface},}\ }\href {\doibase
  10.1063/1.1707989} {\bibfield  {journal} {\bibinfo  {journal} {J. Appl.
  Phys.}\ }\textbf {\bibinfo {volume} {37}},\ \bibinfo {pages} {4139--4141}
  (\bibinfo {year} {1966})}\BibitemShut {NoStop}%
\bibitem [{\citenamefont {Kogan}(1994)}]{Kog94prb}%
  \BibitemOpen
  \bibfield  {author} {\bibinfo {author} {\bibfnamefont {V.~G.}\ \bibnamefont
  {Kogan}},\ }\bibfield  {title} {\enquote {\bibinfo {title} {Pearl's vortex
  near the film edge},}\ }\href {\doibase 10.1103/PhysRevB.49.15874} {\bibfield
   {journal} {\bibinfo  {journal} {Phys. Rev. B}\ }\textbf {\bibinfo {volume}
  {49}},\ \bibinfo {pages} {15874--15878} (\bibinfo {year} {1994})}\BibitemShut
  {NoStop}%
\bibitem [{\citenamefont {Mikitik}(2021)}]{Mik21prb}%
  \BibitemOpen
  \bibfield  {author} {\bibinfo {author} {\bibfnamefont {G.~P.}\ \bibnamefont
  {Mikitik}},\ }\bibfield  {title} {\enquote {\bibinfo {title} {Critical
  current in thin flat superconductors with {Bean-Livingston} and geometrical
  barriers},}\ }\href {\doibase 10.1103/PhysRevB.104.094526} {\bibfield
  {journal} {\bibinfo  {journal} {Phys. Rev. B}\ }\textbf {\bibinfo {volume}
  {104}},\ \bibinfo {pages} {094526} (\bibinfo {year} {2021})}\BibitemShut
  {NoStop}%
\bibitem [{\citenamefont {Adami}\ \emph {et~al.}(2013)\citenamefont {Adami},
  \citenamefont {Cerbu}, \citenamefont {Cabosart}, \citenamefont {Motta},
  \citenamefont {Cuppens}, \citenamefont {Ortiz}, \citenamefont {Moshchalkov},
  \citenamefont {Hackens}, \citenamefont {Delamare}, \citenamefont {Van~de
  Vondel},\ and\ \citenamefont {Silhanek}}]{Ada13apl}%
  \BibitemOpen
  \bibfield  {author} {\bibinfo {author} {\bibfnamefont {O.-A.}\ \bibnamefont
  {Adami}}, \bibinfo {author} {\bibfnamefont {D.}~\bibnamefont {Cerbu}},
  \bibinfo {author} {\bibfnamefont {D.}~\bibnamefont {Cabosart}}, \bibinfo
  {author} {\bibfnamefont {M.}~\bibnamefont {Motta}}, \bibinfo {author}
  {\bibfnamefont {J.}~\bibnamefont {Cuppens}}, \bibinfo {author} {\bibfnamefont
  {W.~A.}\ \bibnamefont {Ortiz}}, \bibinfo {author} {\bibfnamefont {V.~V.}\
  \bibnamefont {Moshchalkov}}, \bibinfo {author} {\bibfnamefont
  {B.}~\bibnamefont {Hackens}}, \bibinfo {author} {\bibfnamefont
  {R.}~\bibnamefont {Delamare}}, \bibinfo {author} {\bibfnamefont
  {J.}~\bibnamefont {Van~de Vondel}}, \ and\ \bibinfo {author} {\bibfnamefont
  {A.~V.}\ \bibnamefont {Silhanek}},\ }\bibfield  {title} {\enquote {\bibinfo
  {title} {Current crowding effects in superconducting corner-shaped {Al}
  microstrips},}\ }\href {\doibase http://dx.doi.org/10.1063/1.4790625}
  {\bibfield  {journal} {\bibinfo  {journal} {Appl. Phys. Lett.}\ }\textbf
  {\bibinfo {volume} {102}},\ \bibinfo {eid} {052603} (\bibinfo {year}
  {2013})}\BibitemShut {NoStop}%
\bibitem [{\citenamefont {Niessen}\ and\ \citenamefont
  {Weijsenfeld}(1969)}]{Nie69jap}%
  \BibitemOpen
  \bibfield  {author} {\bibinfo {author} {\bibfnamefont {A.~K.}\ \bibnamefont
  {Niessen}}\ and\ \bibinfo {author} {\bibfnamefont {C.~H.}\ \bibnamefont
  {Weijsenfeld}},\ }\bibfield  {title} {\enquote {\bibinfo {title} {Anisotropic
  pinning and guided motion of vortices in type-{II} superconductors},}\ }\href
  {\doibase 10.1063/1.1657066} {\bibfield  {journal} {\bibinfo  {journal} {J.
  Appl. Phys.}\ }\textbf {\bibinfo {volume} {40}},\ \bibinfo {pages} {384--393}
  (\bibinfo {year} {1969})}\BibitemShut {NoStop}%
\bibitem [{\citenamefont {Silhanek}\ \emph {et~al.}(2003)\citenamefont
  {Silhanek}, \citenamefont {Van~Look}, \citenamefont {Raedts}, \citenamefont
  {Jonckheere},\ and\ \citenamefont {Moshchalkov}}]{Sil03prb}%
  \BibitemOpen
  \bibfield  {author} {\bibinfo {author} {\bibfnamefont {A.~V.}\ \bibnamefont
  {Silhanek}}, \bibinfo {author} {\bibfnamefont {L.}~\bibnamefont {Van~Look}},
  \bibinfo {author} {\bibfnamefont {S.}~\bibnamefont {Raedts}}, \bibinfo
  {author} {\bibfnamefont {R.}~\bibnamefont {Jonckheere}}, \ and\ \bibinfo
  {author} {\bibfnamefont {V.~V.}\ \bibnamefont {Moshchalkov}},\ }\bibfield
  {title} {\enquote {\bibinfo {title} {Guided vortex motion in superconductors
  with a square antidot array},}\ }\href {\doibase 10.1103/PhysRevB.68.214504}
  {\bibfield  {journal} {\bibinfo  {journal} {Phys. Rev. B}\ }\textbf {\bibinfo
  {volume} {68}},\ \bibinfo {pages} {214504} (\bibinfo {year}
  {2003})}\BibitemShut {NoStop}%
\bibitem [{\citenamefont {Shklovskij}\ and\ \citenamefont
  {Dobrovolskiy}(2006)}]{Shk06prb}%
  \BibitemOpen
  \bibfield  {author} {\bibinfo {author} {\bibfnamefont {V.~A.}\ \bibnamefont
  {Shklovskij}}\ and\ \bibinfo {author} {\bibfnamefont {O.~V.}\ \bibnamefont
  {Dobrovolskiy}},\ }\bibfield  {title} {\enquote {\bibinfo {title} {Influence
  of pointlike disorder on the guiding of vortices and the {Hall} effect in a
  washboard planar pinning potential},}\ }\href {\doibase
  10.1103/PhysRevB.74.104511} {\bibfield  {journal} {\bibinfo  {journal} {Phys.
  Rev. B}\ }\textbf {\bibinfo {volume} {74}},\ \bibinfo {pages} {104511--1--14}
  (\bibinfo {year} {2006})}\BibitemShut {NoStop}%
\bibitem [{\citenamefont {Reichhardt}\ and\ \citenamefont
  {Reichhardt}(2008)}]{Rei08prb}%
  \BibitemOpen
  \bibfield  {author} {\bibinfo {author} {\bibfnamefont {C.}~\bibnamefont
  {Reichhardt}}\ and\ \bibinfo {author} {\bibfnamefont {C.~J.~Olson}\
  \bibnamefont {Reichhardt}},\ }\bibfield  {title} {\enquote {\bibinfo {title}
  {Moving vortex phases, dynamical symmetry breaking, and jamming for vortices
  in honeycomb pinning arrays},}\ }\href {\doibase 10.1103/PhysRevB.78.224511}
  {\bibfield  {journal} {\bibinfo  {journal} {Phys. Rev. B}\ }\textbf {\bibinfo
  {volume} {78}},\ \bibinfo {pages} {224511} (\bibinfo {year}
  {2008})}\BibitemShut {NoStop}%
\bibitem [{\citenamefont {W\"ordenweber}\ \emph {et~al.}(2012)\citenamefont
  {W\"ordenweber}, \citenamefont {Hollmann}, \citenamefont {Schubert},
  \citenamefont {Kutzner},\ and\ \citenamefont {Panaitov}}]{Wor12prb}%
  \BibitemOpen
  \bibfield  {author} {\bibinfo {author} {\bibfnamefont {R.}~\bibnamefont
  {W\"ordenweber}}, \bibinfo {author} {\bibfnamefont {E.}~\bibnamefont
  {Hollmann}}, \bibinfo {author} {\bibfnamefont {J.}~\bibnamefont {Schubert}},
  \bibinfo {author} {\bibfnamefont {R.}~\bibnamefont {Kutzner}}, \ and\
  \bibinfo {author} {\bibfnamefont {G.}~\bibnamefont {Panaitov}},\ }\bibfield
  {title} {\enquote {\bibinfo {title} {Regimes of flux transport at microwave
  frequencies in nanostructured high-${T}_{c}$ films},}\ }\href {\doibase
  10.1103/PhysRevB.85.064503} {\bibfield  {journal} {\bibinfo  {journal} {Phys.
  Rev. B}\ }\textbf {\bibinfo {volume} {85}},\ \bibinfo {pages} {064503--1--6}
  (\bibinfo {year} {2012})}\BibitemShut {NoStop}%
\bibitem [{\citenamefont {Dobrovolskiy}\ \emph {et~al.}(2019)\citenamefont
  {Dobrovolskiy}, \citenamefont {Bevz}, \citenamefont {Begun}, \citenamefont
  {Sachser}, \citenamefont {Vovk},\ and\ \citenamefont {Huth}}]{Dob19pra}%
  \BibitemOpen
  \bibfield  {author} {\bibinfo {author} {\bibfnamefont {O.~V.}\ \bibnamefont
  {Dobrovolskiy}}, \bibinfo {author} {\bibfnamefont {V.~M.}\ \bibnamefont
  {Bevz}}, \bibinfo {author} {\bibfnamefont {E.}~\bibnamefont {Begun}},
  \bibinfo {author} {\bibfnamefont {R.}~\bibnamefont {Sachser}}, \bibinfo
  {author} {\bibfnamefont {R.~V.}\ \bibnamefont {Vovk}}, \ and\ \bibinfo
  {author} {\bibfnamefont {M.}~\bibnamefont {Huth}},\ }\bibfield  {title}
  {\enquote {\bibinfo {title} {Fast dynamics of guided magnetic flux quanta},}\
  }\href {\doibase 10.1103/PhysRevApplied.11.054064} {\bibfield  {journal}
  {\bibinfo  {journal} {Phys. Rev. Appl.}\ }\textbf {\bibinfo {volume} {11}},\
  \bibinfo {pages} {054064} (\bibinfo {year} {2019})}\BibitemShut {NoStop}%
\bibitem [{\citenamefont {Vinokur}\ \emph {et~al.}(1993)\citenamefont
  {Vinokur}, \citenamefont {Geshkenbein}, \citenamefont {Feigel'man},\ and\
  \citenamefont {Blatter}}]{Vin93prl}%
  \BibitemOpen
  \bibfield  {author} {\bibinfo {author} {\bibfnamefont {V.~M.}\ \bibnamefont
  {Vinokur}}, \bibinfo {author} {\bibfnamefont {V.~B.}\ \bibnamefont
  {Geshkenbein}}, \bibinfo {author} {\bibfnamefont {M.~V.}\ \bibnamefont
  {Feigel'man}}, \ and\ \bibinfo {author} {\bibfnamefont {G.}~\bibnamefont
  {Blatter}},\ }\bibfield  {title} {\enquote {\bibinfo {title} {Scaling of the
  hall resistivity in high-{Tc} superconductors},}\ }\href {\doibase
  10.1103/PhysRevLett.71.1242} {\bibfield  {journal} {\bibinfo  {journal}
  {Phys. Rev. Lett.}\ }\textbf {\bibinfo {volume} {71}},\ \bibinfo {pages}
  {1242--1245} (\bibinfo {year} {1993})}\BibitemShut {NoStop}%
\bibitem [{\citenamefont {Lang}\ \emph {et~al.}(2001)\citenamefont {Lang},
  \citenamefont {G{\"o}b}, \citenamefont {Pedarnig}, \citenamefont
  {R{\"o}ssler},\ and\ \citenamefont {B{\"a}uerle}}]{Lan01pcs}%
  \BibitemOpen
  \bibfield  {author} {\bibinfo {author} {\bibfnamefont {W.}~\bibnamefont
  {Lang}}, \bibinfo {author} {\bibfnamefont {W.}~\bibnamefont {G{\"o}b}},
  \bibinfo {author} {\bibfnamefont {J.~D.}\ \bibnamefont {Pedarnig}}, \bibinfo
  {author} {\bibfnamefont {R.}~\bibnamefont {R{\"o}ssler}}, \ and\ \bibinfo
  {author} {\bibfnamefont {D.}~\bibnamefont {B{\"a}uerle}},\ }\bibfield
  {title} {\enquote {\bibinfo {title} {Anomalous {Hall} effect and vortex
  pinning in high-{$T_c$} superconductors},}\ }\href
  {https://www.sciencedirect.com/science/article/pii/S0921453401008413}
  {\bibfield  {journal} {\bibinfo  {journal} {Physica C}\ }\textbf {\bibinfo
  {volume} {364-365}},\ \bibinfo {pages} {518--521} (\bibinfo {year}
  {2001})}\BibitemShut {NoStop}%
\bibitem [{\citenamefont {W\"ordenweber}\ \emph {et~al.}(2009)\citenamefont
  {W\"ordenweber}, \citenamefont {Hollmann}, \citenamefont {Schubert},
  \citenamefont {Kutzner},\ and\ \citenamefont {Ghosh}}]{Wor09apl}%
  \BibitemOpen
  \bibfield  {author} {\bibinfo {author} {\bibfnamefont {R.}~\bibnamefont
  {W\"ordenweber}}, \bibinfo {author} {\bibfnamefont {E.}~\bibnamefont
  {Hollmann}}, \bibinfo {author} {\bibfnamefont {J.}~\bibnamefont {Schubert}},
  \bibinfo {author} {\bibfnamefont {R.}~\bibnamefont {Kutzner}}, \ and\
  \bibinfo {author} {\bibfnamefont {A.~K.}\ \bibnamefont {Ghosh}},\ }\bibfield
  {title} {\enquote {\bibinfo {title} {Pattern induced phase transition of
  vortex motion in high-{$T_c$} films},}\ }\href {\doibase
  http://dx.doi.org/10.1063/1.3139077} {\bibfield  {journal} {\bibinfo
  {journal} {Appl. Phys. Lett.}\ }\textbf {\bibinfo {volume} {94}},\ \bibinfo
  {eid} {202501} (\bibinfo {year} {2009})}\BibitemShut {NoStop}%
\bibitem [{\citenamefont {Puica}\ \emph {et~al.}(2009)\citenamefont {Puica},
  \citenamefont {Lang}, \citenamefont {Siraj}, \citenamefont {Pedarnig},\ and\
  \citenamefont {B\"auerle}}]{Pui09prb}%
  \BibitemOpen
  \bibfield  {author} {\bibinfo {author} {\bibfnamefont {I.}~\bibnamefont
  {Puica}}, \bibinfo {author} {\bibfnamefont {W.}~\bibnamefont {Lang}},
  \bibinfo {author} {\bibfnamefont {K.}~\bibnamefont {Siraj}}, \bibinfo
  {author} {\bibfnamefont {J.~D.}\ \bibnamefont {Pedarnig}}, \ and\ \bibinfo
  {author} {\bibfnamefont {D.}~\bibnamefont {B\"auerle}},\ }\bibfield  {title}
  {\enquote {\bibinfo {title} {Non-ohmic {Hall} resistivity observed above the
  critical temperature in the high-temperature superconductor
  {YBa$_2$Cu$_3$O$_{7-\delta}$}},}\ }\href {\doibase
  10.1103/PhysRevB.79.094522} {\bibfield  {journal} {\bibinfo  {journal} {Phys.
  Rev. B}\ }\textbf {\bibinfo {volume} {79}},\ \bibinfo {pages} {094522}
  (\bibinfo {year} {2009})}\BibitemShut {NoStop}%
\bibitem [{\citenamefont {Zechner}\ \emph {et~al.}(2018)\citenamefont
  {Zechner}, \citenamefont {Lang}, \citenamefont {Dosmailov}, \citenamefont
  {Bodea},\ and\ \citenamefont {Pedarnig}}]{Zec18prb}%
  \BibitemOpen
  \bibfield  {author} {\bibinfo {author} {\bibfnamefont {G.}~\bibnamefont
  {Zechner}}, \bibinfo {author} {\bibfnamefont {W.}~\bibnamefont {Lang}},
  \bibinfo {author} {\bibfnamefont {M.}~\bibnamefont {Dosmailov}}, \bibinfo
  {author} {\bibfnamefont {M.~A.}\ \bibnamefont {Bodea}}, \ and\ \bibinfo
  {author} {\bibfnamefont {J.~D.}\ \bibnamefont {Pedarnig}},\ }\bibfield
  {title} {\enquote {\bibinfo {title} {Transverse vortex commensurability
  effect and sign change of the {Hall} voltage in superconducting
  {${\mathrm{YBa}}_{2}{\mathrm{Cu}}_{3}{\mathrm{O}}_{7\ensuremath{-}\ensuremath{\delta}}$
  } thin films with a nanoscale periodic pinning landscape},}\ }\href {\doibase
  10.1103/PhysRevB.98.104508} {\bibfield  {journal} {\bibinfo  {journal} {Phys.
  Rev. B}\ }\textbf {\bibinfo {volume} {98}},\ \bibinfo {pages} {104508}
  (\bibinfo {year} {2018})}\BibitemShut {NoStop}%
\bibitem [{\citenamefont {Richter}\ \emph {et~al.}(2021)\citenamefont
  {Richter}, \citenamefont {Lang}, \citenamefont {Peruzzi}, \citenamefont
  {Hattmansdorfer}, \citenamefont {Durrell},\ and\ \citenamefont
  {Pedarnig}}]{Ric21sst}%
  \BibitemOpen
  \bibfield  {author} {\bibinfo {author} {\bibfnamefont {H.}~\bibnamefont
  {Richter}}, \bibinfo {author} {\bibfnamefont {W.}~\bibnamefont {Lang}},
  \bibinfo {author} {\bibfnamefont {M.}~\bibnamefont {Peruzzi}}, \bibinfo
  {author} {\bibfnamefont {H.}~\bibnamefont {Hattmansdorfer}}, \bibinfo
  {author} {\bibfnamefont {J.~H.}\ \bibnamefont {Durrell}}, \ and\ \bibinfo
  {author} {\bibfnamefont {J.~D.}\ \bibnamefont {Pedarnig}},\ }\bibfield
  {title} {\enquote {\bibinfo {title} {Resistivity, {Hall} effect, and
  anisotropic superconducting coherence lengths of {HgBa$_2$CaCu$_2$O$_6$} thin
  films with different morphology},}\ }\href {\doibase
  10.1088/1361-6668/abdedf} {\bibfield  {journal} {\bibinfo  {journal}
  {Supercond. Sci. Technol.}\ }\textbf {\bibinfo {volume} {34}},\ \bibinfo
  {pages} {035031} (\bibinfo {year} {2021})}\BibitemShut {NoStop}%
\bibitem [{\citenamefont {Glazman}(1986)}]{Gla86ltp}%
  \BibitemOpen
  \bibfield  {author} {\bibinfo {author} {\bibfnamefont {L.~I.}\ \bibnamefont
  {Glazman}},\ }\bibfield  {title} {\enquote {\bibinfo {title} {Vortex induced
  transverse voltage within film},}\ }\href@noop {} {\bibfield  {journal}
  {\bibinfo  {journal} {Sov. J. Low Temp. Phys.}\ }\textbf {\bibinfo {volume}
  {12}},\ \bibinfo {pages} {389} (\bibinfo {year} {1986})}\BibitemShut
  {NoStop}%
\bibitem [{\citenamefont {Antonova}\ \emph {et~al.}(1991)\citenamefont
  {Antonova}, \citenamefont {Zakosarenko}, \citenamefont {Il'ichev},
  \citenamefont {Kuznetsov},\ and\ \citenamefont {Tulin}}]{Ant91etp}%
  \BibitemOpen
  \bibfield  {author} {\bibinfo {author} {\bibfnamefont {I.~Yu.}\ \bibnamefont
  {Antonova}}, \bibinfo {author} {\bibfnamefont {V.~M.}\ \bibnamefont
  {Zakosarenko}}, \bibinfo {author} {\bibfnamefont {E.V.}\ \bibnamefont
  {Il'ichev}}, \bibinfo {author} {\bibfnamefont {V.~I.}\ \bibnamefont
  {Kuznetsov}}, \ and\ \bibinfo {author} {\bibfnamefont {V.~A.}\ \bibnamefont
  {Tulin}},\ }\bibfield  {title} {\enquote {\bibinfo {title} {Observation of a
  nonmonotonic transverse voltage induced by vortex motion in a supercondcuting
  thin film},}\ }\href@noop {} {\bibfield  {journal} {\bibinfo  {journal} {JETP
  Lett.}\ }\textbf {\bibinfo {volume} {54}},\ \bibinfo {pages} {505} (\bibinfo
  {year} {1991})}\BibitemShut {NoStop}%
\bibitem [{\citenamefont {Pearl}(1964)}]{Pea64apl}%
  \BibitemOpen
  \bibfield  {author} {\bibinfo {author} {\bibfnamefont {J.}~\bibnamefont
  {Pearl}},\ }\bibfield  {title} {\enquote {\bibinfo {title} {Current
  distribution in superconducting films carrying quantized fluxoids},}\ }\href
  {\doibase 10.1063/1.1754056} {\bibfield  {journal} {\bibinfo  {journal}
  {Appl. Phys. Lett.}\ }\textbf {\bibinfo {volume} {5}},\ \bibinfo {pages}
  {65--66} (\bibinfo {year} {1964})}\BibitemShut {NoStop}%
\bibitem [{\citenamefont {de~Gennes}(1966)}]{Gen66boo}%
  \BibitemOpen
  \bibfield  {author} {\bibinfo {author} {\bibfnamefont {P.~G.}\ \bibnamefont
  {de~Gennes}},\ }\href@noop {} {\emph {\bibinfo {title} {Superconductivity of
  Metals and Alloys}}}\ (\bibinfo  {publisher} {Benjamin, New York},\ \bibinfo
  {year} {1966})\BibitemShut {NoStop}%
\bibitem [{\citenamefont {Blatter}\ \emph {et~al.}(1994)\citenamefont
  {Blatter}, \citenamefont {Feigel'man}, \citenamefont {Geshkenbein},
  \citenamefont {Larkin},\ and\ \citenamefont {Vinokur}}]{Bla94rmp}%
  \BibitemOpen
  \bibfield  {author} {\bibinfo {author} {\bibfnamefont {G.}~\bibnamefont
  {Blatter}}, \bibinfo {author} {\bibfnamefont {M.~V.}\ \bibnamefont
  {Feigel'man}}, \bibinfo {author} {\bibfnamefont {V.~B.}\ \bibnamefont
  {Geshkenbein}}, \bibinfo {author} {\bibfnamefont {A.~I.}\ \bibnamefont
  {Larkin}}, \ and\ \bibinfo {author} {\bibfnamefont {V.~M.}\ \bibnamefont
  {Vinokur}},\ }\bibfield  {title} {\enquote {\bibinfo {title} {Vortices in
  high-temperature superconductors},}\ }\href {\doibase
  10.1103/RevModPhys.66.1125} {\bibfield  {journal} {\bibinfo  {journal} {Rev.
  Mod. Phys.}\ }\textbf {\bibinfo {volume} {66}},\ \bibinfo {pages}
  {1125--1388} (\bibinfo {year} {1994})}\BibitemShut {NoStop}%
\bibitem [{\citenamefont {Leo}\ \emph {et~al.}(2020)\citenamefont {Leo},
  \citenamefont {Nigro}, \citenamefont {Braccini}, \citenamefont {Sylva},
  \citenamefont {Provino}, \citenamefont {Galluzzi}, \citenamefont
  {Polichetti}, \citenamefont {Ferdeghini}, \citenamefont {Putti},\ and\
  \citenamefont {Grimaldi}}]{Leo20sst}%
  \BibitemOpen
  \bibfield  {author} {\bibinfo {author} {\bibfnamefont {A.}~\bibnamefont
  {Leo}}, \bibinfo {author} {\bibfnamefont {A.}~\bibnamefont {Nigro}}, \bibinfo
  {author} {\bibfnamefont {V.}~\bibnamefont {Braccini}}, \bibinfo {author}
  {\bibfnamefont {G.}~\bibnamefont {Sylva}}, \bibinfo {author} {\bibfnamefont
  {A.}~\bibnamefont {Provino}}, \bibinfo {author} {\bibfnamefont
  {A.}~\bibnamefont {Galluzzi}}, \bibinfo {author} {\bibfnamefont
  {M.}~\bibnamefont {Polichetti}}, \bibinfo {author} {\bibfnamefont
  {C.}~\bibnamefont {Ferdeghini}}, \bibinfo {author} {\bibfnamefont
  {M.}~\bibnamefont {Putti}}, \ and\ \bibinfo {author} {\bibfnamefont
  {G.}~\bibnamefont {Grimaldi}},\ }\bibfield  {title} {\enquote {\bibinfo
  {title} {Flux flow instability as a probe for quasiparticle energy relaxation
  time in {Fe}-chalcogenides},}\ }\href {\doibase 10.1088/1361-6668/abaec1}
  {\bibfield  {journal} {\bibinfo  {journal} {Supercond. Sci. Technol.}\
  }\textbf {\bibinfo {volume} {33}},\ \bibinfo {pages} {104005} (\bibinfo
  {year} {2020})}\BibitemShut {NoStop}%
\bibitem [{\citenamefont {Ustavschikov}\ \emph {et~al.}(2020)\citenamefont
  {Ustavschikov}, \citenamefont {Levichev}, \citenamefont {Pashenkin},
  \citenamefont {Klushin},\ and\ \citenamefont {Vodolazov}}]{Ust20sst}%
  \BibitemOpen
  \bibfield  {author} {\bibinfo {author} {\bibfnamefont {S.~S.}\ \bibnamefont
  {Ustavschikov}}, \bibinfo {author} {\bibfnamefont {M.~Yu.}\ \bibnamefont
  {Levichev}}, \bibinfo {author} {\bibfnamefont {I.~Yu.}\ \bibnamefont
  {Pashenkin}}, \bibinfo {author} {\bibfnamefont {A.~M.}\ \bibnamefont
  {Klushin}}, \ and\ \bibinfo {author} {\bibfnamefont {D.~Yu.}\ \bibnamefont
  {Vodolazov}},\ }\bibfield  {title} {\enquote {\bibinfo {title} {Approaching
  depairing current in dirty thin superconducting strip covered by low
  resistive normal metal},}\ }\href {\doibase 10.1088/1361-6668/abc2ad}
  {\bibfield  {journal} {\bibinfo  {journal} {Supercond. Sci. Technol.}\
  }\textbf {\bibinfo {volume} {34}},\ \bibinfo {pages} {015004} (\bibinfo
  {year} {2020})}\BibitemShut {NoStop}%
\bibitem [{\citenamefont {Pathirana}\ and\ \citenamefont
  {Gurevich}(2020)}]{Pat20prb}%
  \BibitemOpen
  \bibfield  {author} {\bibinfo {author} {\bibfnamefont {W.~P. M.~R.}\
  \bibnamefont {Pathirana}}\ and\ \bibinfo {author} {\bibfnamefont
  {A.}~\bibnamefont {Gurevich}},\ }\bibfield  {title} {\enquote {\bibinfo
  {title} {Nonlinear dynamics and dissipation of a curvilinear vortex driven by
  a strong time-dependent {Meissner} current},}\ }\href {\doibase
  10.1103/PhysRevB.101.064504} {\bibfield  {journal} {\bibinfo  {journal}
  {Phys. Rev. B}\ }\textbf {\bibinfo {volume} {101}},\ \bibinfo {pages}
  {064504} (\bibinfo {year} {2020})}\BibitemShut {NoStop}%
\bibitem [{\citenamefont {Rhoderick}\ and\ \citenamefont
  {Wilson}(1962)}]{Rho62nat}%
  \BibitemOpen
  \bibfield  {author} {\bibinfo {author} {\bibfnamefont {E.~H.}\ \bibnamefont
  {Rhoderick}}\ and\ \bibinfo {author} {\bibfnamefont {E.~M.}\ \bibnamefont
  {Wilson}},\ }\bibfield  {title} {\enquote {\bibinfo {title} {Current
  distribution in thin superconducting films},}\ }\href {\doibase
  10.1038/1941167b0} {\bibfield  {journal} {\bibinfo  {journal} {Nature}\
  }\textbf {\bibinfo {volume} {194}},\ \bibinfo {pages} {1167--1168} (\bibinfo
  {year} {1962})}\BibitemShut {NoStop}%
\bibitem [{\citenamefont {Larkin}\ and\ \citenamefont
  {Ovchinnikov}(1971)}]{Lar71etp}%
  \BibitemOpen
  \bibfield  {author} {\bibinfo {author} {\bibfnamefont {A.I.}\ \bibnamefont
  {Larkin}}\ and\ \bibinfo {author} {\bibfnamefont {Yu.N.}\ \bibnamefont
  {Ovchinnikov}},\ }\bibfield  {title} {\enquote {\bibinfo {title} {Influence
  of inhomogeneities on superconductor properties.}}\ }\href@noop {} {\bibfield
   {journal} {\bibinfo  {journal} {Sov. Phys. JETP}\ }\textbf {\bibinfo
  {volume} {34}},\ \bibinfo {pages} {651} (\bibinfo {year} {1971})}\BibitemShut
  {NoStop}%
\bibitem [{\citenamefont {Tinkham}(2004)}]{Tin04boo}%
  \BibitemOpen
  \bibfield  {author} {\bibinfo {author} {\bibfnamefont {M.}~\bibnamefont
  {Tinkham}},\ }\href@noop {} {\emph {\bibinfo {title} {Introduction to
  Superconductivity}}}\ (\bibinfo  {publisher} {Mineola, New York},\ \bibinfo
  {year} {2004})\BibitemShut {NoStop}%
\bibitem [{\citenamefont {Larkin}\ and\ \citenamefont
  {Ovchinnikov}(1975)}]{Lar75etp}%
  \BibitemOpen
  \bibfield  {author} {\bibinfo {author} {\bibfnamefont {A.~I.}\ \bibnamefont
  {Larkin}}\ and\ \bibinfo {author} {\bibfnamefont {Yu.~N.}\ \bibnamefont
  {Ovchinnikov}},\ }\bibfield  {title} {\enquote {\bibinfo {title} {Nonlinear
  conductivity of superconductors in the mixed state},}\ }\href
  {http://jetp.ras.ru/cgi-bin/e/index/e/41/5/p960?a=list} {\bibfield  {journal}
  {\bibinfo  {journal} {J. Exp. Theor. Phys.}\ }\textbf {\bibinfo {volume}
  {41}},\ \bibinfo {pages} {960} (\bibinfo {year} {1975})}\BibitemShut
  {NoStop}%
\bibitem [{\citenamefont {Larkin}\ and\ \citenamefont
  {Ovchinnikov}(1986)}]{Lar86inb}%
  \BibitemOpen
  \bibfield  {author} {\bibinfo {author} {\bibfnamefont {A.~I.}\ \bibnamefont
  {Larkin}}\ and\ \bibinfo {author} {\bibfnamefont {Y.~N.}\ \bibnamefont
  {Ovchinnikov}},\ }\enquote {\bibinfo {title} {Nonequilibrium
  superconductivity},}\ \ (\bibinfo  {publisher} {Elsevier, Amsterdam},\
  \bibinfo {year} {1986})\ p.\ \bibinfo {pages} {493}\BibitemShut {NoStop}%
\bibitem [{\citenamefont {Bezuglyj}\ and\ \citenamefont
  {Shklovskij}(1992)}]{Bez92pcs}%
  \BibitemOpen
  \bibfield  {author} {\bibinfo {author} {\bibfnamefont {A.I.}\ \bibnamefont
  {Bezuglyj}}\ and\ \bibinfo {author} {\bibfnamefont {V.A.}\ \bibnamefont
  {Shklovskij}},\ }\bibfield  {title} {\enquote {\bibinfo {title} {Effect of
  self-heating on flux flow instability in a superconductor near {$T_c$}},}\
  }\href {\doibase 10.1016/0921-4534(92)90165-9} {\bibfield  {journal}
  {\bibinfo  {journal} {Physica C}\ }\textbf {\bibinfo {volume} {202}},\
  \bibinfo {pages} {234} (\bibinfo {year} {1992})}\BibitemShut {NoStop}%
\bibitem [{\citenamefont {Romijn}\ \emph {et~al.}(1982)\citenamefont {Romijn},
  \citenamefont {Klapwijk}, \citenamefont {Renne},\ and\ \citenamefont
  {Mooij}}]{Rom82prb}%
  \BibitemOpen
  \bibfield  {author} {\bibinfo {author} {\bibfnamefont {J.}~\bibnamefont
  {Romijn}}, \bibinfo {author} {\bibfnamefont {T.~M.}\ \bibnamefont
  {Klapwijk}}, \bibinfo {author} {\bibfnamefont {M.~J.}\ \bibnamefont {Renne}},
  \ and\ \bibinfo {author} {\bibfnamefont {J.~E.}\ \bibnamefont {Mooij}},\
  }\bibfield  {title} {\enquote {\bibinfo {title} {Critical pair-breaking
  current in superconducting aluminum strips far below ${T}_{c}$},}\ }\href
  {\doibase 10.1103/PhysRevB.26.3648} {\bibfield  {journal} {\bibinfo
  {journal} {Phys. Rev. B}\ }\textbf {\bibinfo {volume} {26}},\ \bibinfo
  {pages} {3648--3655} (\bibinfo {year} {1982})}\BibitemShut {NoStop}%
\bibitem [{\citenamefont {Clem}\ and\ \citenamefont {Kogan}(2012)}]{Cle12prb}%
  \BibitemOpen
  \bibfield  {author} {\bibinfo {author} {\bibfnamefont {J.~R.}\ \bibnamefont
  {Clem}}\ and\ \bibinfo {author} {\bibfnamefont {V.~G.}\ \bibnamefont
  {Kogan}},\ }\bibfield  {title} {\enquote {\bibinfo {title} {Kinetic impedance
  and depairing in thin and narrow superconducting films},}\ }\href {\doibase
  10.1103/PhysRevB.86.174521} {\bibfield  {journal} {\bibinfo  {journal} {Phys.
  Rev. B}\ }\textbf {\bibinfo {volume} {86}},\ \bibinfo {pages} {174521}
  (\bibinfo {year} {2012})}\BibitemShut {NoStop}%
\bibitem [{\citenamefont {Aranson}\ \emph {et~al.}(2001)\citenamefont
  {Aranson}, \citenamefont {Gurevich},\ and\ \citenamefont
  {Vinokur}}]{Ara01prl}%
  \BibitemOpen
  \bibfield  {author} {\bibinfo {author} {\bibfnamefont {I.}~\bibnamefont
  {Aranson}}, \bibinfo {author} {\bibfnamefont {A.}~\bibnamefont {Gurevich}}, \
  and\ \bibinfo {author} {\bibfnamefont {V.}~\bibnamefont {Vinokur}},\
  }\bibfield  {title} {\enquote {\bibinfo {title} {Vortex avalanches and
  magnetic flux fragmentation in superconductors},}\ }\href {\doibase
  10.1103/PhysRevLett.87.067003} {\bibfield  {journal} {\bibinfo  {journal}
  {Phys. Rev. Lett.}\ }\textbf {\bibinfo {volume} {87}},\ \bibinfo {pages}
  {067003} (\bibinfo {year} {2001})}\BibitemShut {NoStop}%
\bibitem [{\citenamefont {Altshuler}\ and\ \citenamefont
  {Johansen}(2004)}]{Alt04rmp}%
  \BibitemOpen
  \bibfield  {author} {\bibinfo {author} {\bibfnamefont {E.}~\bibnamefont
  {Altshuler}}\ and\ \bibinfo {author} {\bibfnamefont {T.~H.}\ \bibnamefont
  {Johansen}},\ }\bibfield  {title} {\enquote {\bibinfo {title} {Colloquium:
  Experiments in vortex avalanches},}\ }\href {\doibase
  10.1103/RevModPhys.76.471} {\bibfield  {journal} {\bibinfo  {journal} {Rev.
  Mod. Phys.}\ }\textbf {\bibinfo {volume} {76}},\ \bibinfo {pages} {471--487}
  (\bibinfo {year} {2004})}\BibitemShut {NoStop}%
\bibitem [{\citenamefont {Aranson}\ \emph {et~al.}(2005)\citenamefont
  {Aranson}, \citenamefont {Gurevich}, \citenamefont {Welling}, \citenamefont
  {Wijngaarden}, \citenamefont {Vlasko-Vlasov}, \citenamefont {Vinokur},\ and\
  \citenamefont {Welp}}]{Ara05prl}%
  \BibitemOpen
  \bibfield  {author} {\bibinfo {author} {\bibfnamefont {I.~S.}\ \bibnamefont
  {Aranson}}, \bibinfo {author} {\bibfnamefont {A.}~\bibnamefont {Gurevich}},
  \bibinfo {author} {\bibfnamefont {M.~S.}\ \bibnamefont {Welling}}, \bibinfo
  {author} {\bibfnamefont {R.~J.}\ \bibnamefont {Wijngaarden}}, \bibinfo
  {author} {\bibfnamefont {V.~K.}\ \bibnamefont {Vlasko-Vlasov}}, \bibinfo
  {author} {\bibfnamefont {V.~M.}\ \bibnamefont {Vinokur}}, \ and\ \bibinfo
  {author} {\bibfnamefont {U.}~\bibnamefont {Welp}},\ }\bibfield  {title}
  {\enquote {\bibinfo {title} {Dendritic flux avalanches and nonlocal
  electrodynamics in thin superconducting films},}\ }\href {\doibase
  10.1103/PhysRevLett.94.037002} {\bibfield  {journal} {\bibinfo  {journal}
  {Phys. Rev. Lett.}\ }\textbf {\bibinfo {volume} {94}},\ \bibinfo {pages}
  {037002} (\bibinfo {year} {2005})}\BibitemShut {NoStop}%
\bibitem [{\citenamefont {Brisbois}\ \emph {et~al.}(2016)\citenamefont
  {Brisbois}, \citenamefont {Adami}, \citenamefont {Avila}, \citenamefont
  {Motta}, \citenamefont {Ortiz}, \citenamefont {Nguyen}, \citenamefont
  {Vanderbemden}, \citenamefont {Vanderheyden}, \citenamefont {Kramer},\ and\
  \citenamefont {Silhanek}}]{Bri16prb}%
  \BibitemOpen
  \bibfield  {author} {\bibinfo {author} {\bibfnamefont {J.}~\bibnamefont
  {Brisbois}}, \bibinfo {author} {\bibfnamefont {O.-A.}\ \bibnamefont {Adami}},
  \bibinfo {author} {\bibfnamefont {J.~I.}\ \bibnamefont {Avila}}, \bibinfo
  {author} {\bibfnamefont {M.}~\bibnamefont {Motta}}, \bibinfo {author}
  {\bibfnamefont {W.~A.}\ \bibnamefont {Ortiz}}, \bibinfo {author}
  {\bibfnamefont {N.~D.}\ \bibnamefont {Nguyen}}, \bibinfo {author}
  {\bibfnamefont {P.}~\bibnamefont {Vanderbemden}}, \bibinfo {author}
  {\bibfnamefont {B.}~\bibnamefont {Vanderheyden}}, \bibinfo {author}
  {\bibfnamefont {R.~B.~G.}\ \bibnamefont {Kramer}}, \ and\ \bibinfo {author}
  {\bibfnamefont {A.~V.}\ \bibnamefont {Silhanek}},\ }\bibfield  {title}
  {\enquote {\bibinfo {title} {Magnetic flux penetration in nb superconducting
  films with lithographically defined microindentations},}\ }\href {\doibase
  10.1103/PhysRevB.93.054521} {\bibfield  {journal} {\bibinfo  {journal} {Phys.
  Rev. B}\ }\textbf {\bibinfo {volume} {93}},\ \bibinfo {pages} {054521}
  (\bibinfo {year} {2016})}\BibitemShut {NoStop}%
\bibitem [{\citenamefont {Brisbois}\ \emph {et~al.}(2017)\citenamefont
  {Brisbois}, \citenamefont {Gladilin}, \citenamefont {Tempere}, \citenamefont
  {Devreese}, \citenamefont {Moshchalkov}, \citenamefont {Colauto},
  \citenamefont {Motta}, \citenamefont {Johansen}, \citenamefont {Fritzsche},
  \citenamefont {Adami}, \citenamefont {Nguyen}, \citenamefont {Ortiz},
  \citenamefont {Kramer},\ and\ \citenamefont {Silhanek}}]{Bri17prb}%
  \BibitemOpen
  \bibfield  {author} {\bibinfo {author} {\bibfnamefont {J.}~\bibnamefont
  {Brisbois}}, \bibinfo {author} {\bibfnamefont {V.~N.}\ \bibnamefont
  {Gladilin}}, \bibinfo {author} {\bibfnamefont {J.}~\bibnamefont {Tempere}},
  \bibinfo {author} {\bibfnamefont {J.~T.}\ \bibnamefont {Devreese}}, \bibinfo
  {author} {\bibfnamefont {V.~V.}\ \bibnamefont {Moshchalkov}}, \bibinfo
  {author} {\bibfnamefont {F.}~\bibnamefont {Colauto}}, \bibinfo {author}
  {\bibfnamefont {M.}~\bibnamefont {Motta}}, \bibinfo {author} {\bibfnamefont
  {T.~H.}\ \bibnamefont {Johansen}}, \bibinfo {author} {\bibfnamefont
  {J.}~\bibnamefont {Fritzsche}}, \bibinfo {author} {\bibfnamefont {O.-A.}\
  \bibnamefont {Adami}}, \bibinfo {author} {\bibfnamefont {N.~D.}\ \bibnamefont
  {Nguyen}}, \bibinfo {author} {\bibfnamefont {W.~A.}\ \bibnamefont {Ortiz}},
  \bibinfo {author} {\bibfnamefont {R.~B.~G.}\ \bibnamefont {Kramer}}, \ and\
  \bibinfo {author} {\bibfnamefont {A.~V.}\ \bibnamefont {Silhanek}},\
  }\bibfield  {title} {\enquote {\bibinfo {title} {Flux penetration in a
  superconducting film partially capped with a conducting layer},}\ }\href
  {\doibase 10.1103/PhysRevB.95.094506} {\bibfield  {journal} {\bibinfo
  {journal} {Phys. Rev. B}\ }\textbf {\bibinfo {volume} {95}},\ \bibinfo
  {pages} {094506} (\bibinfo {year} {2017})}\BibitemShut {NoStop}%
\bibitem [{\citenamefont {Shaw}\ \emph {et~al.}(2019)\citenamefont {Shaw},
  \citenamefont {Blanco}, \citenamefont {Brisbois}, \citenamefont {Burger},
  \citenamefont {Pinheiro}, \citenamefont {Kramer}, \citenamefont {Motta},
  \citenamefont {Fleury-Frenette}, \citenamefont {Ortiz}, \citenamefont
  {Vanderheyden},\ and\ \citenamefont {Silhanek}}]{Sha19met}%
  \BibitemOpen
  \bibfield  {author} {\bibinfo {author} {\bibfnamefont {G.}~\bibnamefont
  {Shaw}}, \bibinfo {author} {\bibfnamefont {A.S.}\ \bibnamefont {Blanco}},
  \bibinfo {author} {\bibfnamefont {J.}~\bibnamefont {Brisbois}}, \bibinfo
  {author} {\bibfnamefont {L.}~\bibnamefont {Burger}}, \bibinfo {author}
  {\bibfnamefont {L.B.L.G.}\ \bibnamefont {Pinheiro}}, \bibinfo {author}
  {\bibfnamefont {R.B.G.}\ \bibnamefont {Kramer}}, \bibinfo {author}
  {\bibfnamefont {M.}~\bibnamefont {Motta}}, \bibinfo {author} {\bibfnamefont
  {K.}~\bibnamefont {Fleury-Frenette}}, \bibinfo {author} {\bibfnamefont
  {W.A.}\ \bibnamefont {Ortiz}}, \bibinfo {author} {\bibfnamefont
  {B.}~\bibnamefont {Vanderheyden}}, \ and\ \bibinfo {author} {\bibfnamefont
  {A.V.}\ \bibnamefont {Silhanek}},\ }\bibfield  {title} {\enquote {\bibinfo
  {title} {Magnetic recording of superconducting states},}\ }\href {\doibase
  https://doi.org/10.3390/met9101022} {\bibfield  {journal} {\bibinfo
  {journal} {Metals}\ }\textbf {\bibinfo {volume} {9}},\ \bibinfo {pages}
  {1022} (\bibinfo {year} {2019})}\BibitemShut {NoStop}%
\bibitem [{\citenamefont {Jiang}\ \emph {et~al.}(2020)\citenamefont {Jiang},
  \citenamefont {Xue}, \citenamefont {Burger}, \citenamefont {Vanderheyden},
  \citenamefont {Silhanek},\ and\ \citenamefont {Zhou}}]{Jia20prb}%
  \BibitemOpen
  \bibfield  {author} {\bibinfo {author} {\bibfnamefont {Lu}~\bibnamefont
  {Jiang}}, \bibinfo {author} {\bibfnamefont {Cun}\ \bibnamefont {Xue}},
  \bibinfo {author} {\bibfnamefont {L.}~\bibnamefont {Burger}}, \bibinfo
  {author} {\bibfnamefont {B.}~\bibnamefont {Vanderheyden}}, \bibinfo {author}
  {\bibfnamefont {A.~V.}\ \bibnamefont {Silhanek}}, \ and\ \bibinfo {author}
  {\bibfnamefont {You-He}\ \bibnamefont {Zhou}},\ }\bibfield  {title} {\enquote
  {\bibinfo {title} {Selective triggering of magnetic flux avalanches by an
  edge indentation},}\ }\href {\doibase 10.1103/PhysRevB.101.224505} {\bibfield
   {journal} {\bibinfo  {journal} {Phys. Rev. B}\ }\textbf {\bibinfo {volume}
  {101}},\ \bibinfo {pages} {224505} (\bibinfo {year} {2020})}\BibitemShut
  {NoStop}%
\bibitem [{\citenamefont {Colauto}\ \emph {et~al.}(2021)\citenamefont
  {Colauto}, \citenamefont {Motta},\ and\ \citenamefont {Ortiz}}]{Col21sst}%
  \BibitemOpen
  \bibfield  {author} {\bibinfo {author} {\bibfnamefont {F.}~\bibnamefont
  {Colauto}}, \bibinfo {author} {\bibfnamefont {M.}~\bibnamefont {Motta}}, \
  and\ \bibinfo {author} {\bibfnamefont {W.~A.}\ \bibnamefont {Ortiz}},\
  }\bibfield  {title} {\enquote {\bibinfo {title} {Controlling magnetic flux
  penetration in {low-Tc} superconducting films and hybrids},}\ }\href
  {\doibase 10.1088/1361-6668/abac1e} {\bibfield  {journal} {\bibinfo
  {journal} {Supercond. Sci. Technol.}\ }\textbf {\bibinfo {volume} {34}},\
  \bibinfo {pages} {013002} (\bibinfo {year} {2021})}\BibitemShut {NoStop}%
\bibitem [{\citenamefont {Grimaldi}\ \emph {et~al.}(2010)\citenamefont
  {Grimaldi}, \citenamefont {Leo}, \citenamefont {Zola}, \citenamefont {Nigro},
  \citenamefont {Pace}, \citenamefont {Laviano},\ and\ \citenamefont
  {Mezzetti}}]{Gri10prb}%
  \BibitemOpen
  \bibfield  {author} {\bibinfo {author} {\bibfnamefont {G.}~\bibnamefont
  {Grimaldi}}, \bibinfo {author} {\bibfnamefont {A.}~\bibnamefont {Leo}},
  \bibinfo {author} {\bibfnamefont {D.}~\bibnamefont {Zola}}, \bibinfo {author}
  {\bibfnamefont {A.}~\bibnamefont {Nigro}}, \bibinfo {author} {\bibfnamefont
  {S.}~\bibnamefont {Pace}}, \bibinfo {author} {\bibfnamefont {F.}~\bibnamefont
  {Laviano}}, \ and\ \bibinfo {author} {\bibfnamefont {E.}~\bibnamefont
  {Mezzetti}},\ }\bibfield  {title} {\enquote {\bibinfo {title} {Evidence for
  low-field crossover in the vortex critical velocity of type-{II}
  superconducting thin films},}\ }\href {\doibase 10.1103/PhysRevB.82.024512}
  {\bibfield  {journal} {\bibinfo  {journal} {Phys. Rev. B}\ }\textbf {\bibinfo
  {volume} {82}},\ \bibinfo {pages} {024512} (\bibinfo {year}
  {2010})}\BibitemShut {NoStop}%
\bibitem [{\citenamefont {Bezuglyj}\ \emph {et~al.}(2019)\citenamefont
  {Bezuglyj}, \citenamefont {Shklovskij}, \citenamefont {Vovk}, \citenamefont
  {Bevz}, \citenamefont {Huth},\ and\ \citenamefont {Dobrovolskiy}}]{Bez19prb}%
  \BibitemOpen
  \bibfield  {author} {\bibinfo {author} {\bibfnamefont {A.~I.}\ \bibnamefont
  {Bezuglyj}}, \bibinfo {author} {\bibfnamefont {V.~A.}\ \bibnamefont
  {Shklovskij}}, \bibinfo {author} {\bibfnamefont {R.~V.}\ \bibnamefont
  {Vovk}}, \bibinfo {author} {\bibfnamefont {V.~M.}\ \bibnamefont {Bevz}},
  \bibinfo {author} {\bibfnamefont {M.}~\bibnamefont {Huth}}, \ and\ \bibinfo
  {author} {\bibfnamefont {O.~V.}\ \bibnamefont {Dobrovolskiy}},\ }\bibfield
  {title} {\enquote {\bibinfo {title} {Local flux-flow instability in
  superconducting films near {${T}_{c}$}},}\ }\href {\doibase
  10.1103/PhysRevB.99.174518} {\bibfield  {journal} {\bibinfo  {journal} {Phys.
  Rev. B}\ }\textbf {\bibinfo {volume} {99}},\ \bibinfo {pages} {174518}
  (\bibinfo {year} {2019})}\BibitemShut {NoStop}%
\bibitem [{\citenamefont {Vodolazov}\ and\ \citenamefont
  {Peeters}(2007)}]{Vod07prb}%
  \BibitemOpen
  \bibfield  {author} {\bibinfo {author} {\bibfnamefont {D.~Y.}\ \bibnamefont
  {Vodolazov}}\ and\ \bibinfo {author} {\bibfnamefont {F.~M.}\ \bibnamefont
  {Peeters}},\ }\bibfield  {title} {\enquote {\bibinfo {title} {Rearrangement
  of the vortex lattice due to instabilities of vortex flow},}\ }\href
  {\doibase 10.1103/PhysRevB.76.014521} {\bibfield  {journal} {\bibinfo
  {journal} {Phys. Rev. B}\ }\textbf {\bibinfo {volume} {76}},\ \bibinfo
  {pages} {014521} (\bibinfo {year} {2007})}\BibitemShut {NoStop}%
\bibitem [{\citenamefont {Dobrovolskiy}\ \emph {et~al.}(2018)\citenamefont
  {Dobrovolskiy}, \citenamefont {Sachser}, \citenamefont {Huth}, \citenamefont
  {Shklovskij}, \citenamefont {Vovk}, \citenamefont {Bevz},\ and\ \citenamefont
  {Tsindlekht}}]{Dob18apl}%
  \BibitemOpen
  \bibfield  {author} {\bibinfo {author} {\bibfnamefont {O.~V.}\ \bibnamefont
  {Dobrovolskiy}}, \bibinfo {author} {\bibfnamefont {R.}~\bibnamefont
  {Sachser}}, \bibinfo {author} {\bibfnamefont {M.}~\bibnamefont {Huth}},
  \bibinfo {author} {\bibfnamefont {V.~A.}\ \bibnamefont {Shklovskij}},
  \bibinfo {author} {\bibfnamefont {R.~V.}\ \bibnamefont {Vovk}}, \bibinfo
  {author} {\bibfnamefont {V.~M.}\ \bibnamefont {Bevz}}, \ and\ \bibinfo
  {author} {\bibfnamefont {M.~I.}\ \bibnamefont {Tsindlekht}},\ }\bibfield
  {title} {\enquote {\bibinfo {title} {Radiofrequency generation by coherently
  moving fluxons},}\ }\href {\doibase 10.1063/1.5028213} {\bibfield  {journal}
  {\bibinfo  {journal} {Appl. Phys. Lett.}\ }\textbf {\bibinfo {volume}
  {112}},\ \bibinfo {pages} {152601} (\bibinfo {year} {2018})}\BibitemShut
  {NoStop}%
\bibitem [{\citenamefont {Larkin}\ and\ \citenamefont
  {Ovchinnikov}(1976)}]{Lar76etp}%
  \BibitemOpen
  \bibfield  {author} {\bibinfo {author} {\bibfnamefont {A.~I.}\ \bibnamefont
  {Larkin}}\ and\ \bibinfo {author} {\bibfnamefont {Yu.~N.}\ \bibnamefont
  {Ovchinnikov}},\ }\bibfield  {title} {\enquote {\bibinfo {title} {Nonlinear
  conductivity of superconductors in the mixed state},}\ }\href@noop {}
  {\bibfield  {journal} {\bibinfo  {journal} {Sov. Phys. JETP}\ }\textbf
  {\bibinfo {volume} {41}},\ \bibinfo {pages} {960} (\bibinfo {year}
  {1976})}\BibitemShut {NoStop}%
\bibitem [{\citenamefont {Doettinger}\ \emph {et~al.}(1995)\citenamefont
  {Doettinger}, \citenamefont {Huebener},\ and\ \citenamefont
  {K\"uhle}}]{Doe95pcs}%
  \BibitemOpen
  \bibfield  {author} {\bibinfo {author} {\bibfnamefont {S.G.}\ \bibnamefont
  {Doettinger}}, \bibinfo {author} {\bibfnamefont {R.P.}\ \bibnamefont
  {Huebener}}, \ and\ \bibinfo {author} {\bibfnamefont {A.}~\bibnamefont
  {K\"uhle}},\ }\bibfield  {title} {\enquote {\bibinfo {title} {Electronic
  instability during vortex motion in cuprate superconductors regime of low and
  high magnetic fields},}\ }\href {\doibase
  http://dx.doi.org/10.1016/0921-4534(95)00411-4} {\bibfield  {journal}
  {\bibinfo  {journal} {Physica C}\ }\textbf {\bibinfo {volume} {251}},\
  \bibinfo {pages} {285 -- 289} (\bibinfo {year} {1995})}\BibitemShut {NoStop}%
\end{thebibliography}

%

\end{document}